\documentclass[aps,pra,twocolumn,showpacs,superscriptaddress,nofootinbib]{revtex4-1}  

\usepackage{graphicx}  
\usepackage{dcolumn}   
\usepackage{bm}        
\usepackage{amssymb}   
\usepackage{amsmath,epsfig}   
\usepackage{verbatim}  
\usepackage{cases}
\usepackage[usenames,dvipsnames]{xcolor} 
\usepackage{mathrsfs} 

\usepackage{epstopdf}

\definecolor{darkgreen}{rgb}{0.0, 0.5, 0.0}

\newcommand{\ph}{\phantom{0}}
\newcommand{\bml}{\begin{multline}}
\newcommand{\bea}{\begin{eqnarray}}
\newcommand{\eea}{\end{eqnarray}}
\newcommand{\be}{\begin{equation}}
\newcommand{\ee}{\end{equation}}
\newcommand{\bi}{\begin{itemize}}
\newcommand{\ei}{\end{itemize}}

\newcommand{\rr}{\mathbf{r}}
\newcommand{\kk}{{\mathbf{k}}}

\newcommand{\RR}{\mathbf{R}}
\newcommand{\CC}{\mathbf{C}}
\newcommand{\JJ}{\mathbf{J}}

\newcommand{\xx}{\mathbf{x}}
\newcommand{\yy}{\mathbf{y}}
\newcommand{\XX}{\mathbf{X}}
\newcommand{\YY}{\mathbf{Y}}
\newcommand{\vn}{\mathbf{0}}

\newcommand{\gr}{\mathbf{\nabla}}

\newcommand{\Ar}{\mathcal{A}}

\newcommand{\rP}{\mathcal{P}}
\newcommand{\Cr}{\mathcal{C}}

\newcommand{\Sr}{\mathcal{S}}
\newcommand{\Nr}{\mathcal{N}}

\newcommand{\Oo}{\mathbf{\Omega}}
\newcommand{\mA}{m_{\rm A}}
\newcommand{\mB}{m_{\rm B}}

\newcommand{\bcbp}{}

\begin{document}

\title{Three-body recombination in heteronuclear mixtures at finite temperature}

\author{D. S. Petrov}
\affiliation{Universit\'e Paris-Sud, CNRS, LPTMS, UMR8626, Orsay, F-91405, France}

\author{F. Werner}
\affiliation{Laboratoire Kastler Brossel, ENS-PSL, UPMC-Sorbonne Universit\'es, Coll\`ege de France, CNRS, 24 rue Lhomond, 75231 Paris Cedex 05, France}

\date{\today}

\begin{abstract}

Within the universal zero-range theory, we compute the three-body recombination rate to deep molecular states for two identical bosons resonantly interacting with each other and with a third atom of another species, in the absence of weakly bound dimers. The results allow for a quantitative understanding of loss resonances at finite temperature and, combined with experimental data, can be used for testing the Efimov universality and extracting the corresponding three-body parameters in a given system. Curiously, we find that the loss rate can be dramatically enhanced by the resonant heavy-heavy interaction, even for large mass ratios where this interaction is practically irrelevant for the Efimov scaling factor. This effect is important for analysing the recent loss measurements in the Cs-Li mixture.

\end{abstract}
\maketitle

\section{Introduction}

Measuring atomic 
three-body
losses near Feshbach resonances
for large and negative scattering lengths
 has become a major tool for characterizing the Efimov physics in a large variety of ultracold gases,
not only for identical 
bosonic atoms~\cite{RevueBraaten2,EfimovReviewGrimm,KhaykovichEfimovCRAS,GrimmEfimovEvidence,KhaykovichEfimov,KhaykovichEfimov2,Berninger2011uot,jin_contact_BEC,Roy2013tot,UnitaryBoseGas,hadzibabic_unitary_bose,Huang2014oot,Grimm_finite_range_effects,eismann_loss_dynamics}
but also for homonuclear~\cite{Ottenstein2008cso,ohara_excited_trimer_li6,3BP_li6}
 and heteronuclear mixtures~\cite{Florence_KRb,Bloom2013tou,heidelberg_LiCs_efimov,chin_LiCs_efimov}. The peaks in the 
three-body loss rate as a function of magnetic field
 mark passages of Efimov trimers  
through the free-atom scattering threshold. If the size of such a trimer is much smaller than the typical de Broglie wavelengths in the gas, the corresponding peak is most visible and is well described by zero-temperature theory~\cite{RevueBraaten2}. Then, according to the Efimov discrete scaling invariance, the next peak is expected to occur when the two-body scattering length is multiplied by the so-called Efimov period, the trimer being proportionally larger. The Efimov period is numerically quite big, so that when trying to observe multiple successive peaks, i.e., trying to test the Efimov scenario, one very soon faces the problem that Efimov states become too large and the loss rate saturates to a constant value~\cite{UnitaryBoseGas,hadzibabic_unitary_bose,eismann_loss_dynamics,DIncao2004lou}. 

For identical bosons the Efimov period equals 22.7~\cite{Efimov} and two loss peaks separated by approximately this factor have recently
been observed in Cs~\cite{Huang2014oot,Grimm_finite_range_effects}. However, the second (excited-state) peak is already close to the saturation regime and its quantitative characterization has been done relying on the finite-temperature theory developed in Ref.~\cite{UnitaryBoseGas} for three identical bosons. Similarly, an Efimov loss feature 
for large negative scattering lengths
in the mixture of three hyperfine states of $^6{\rm Li}$ has been recently reanalyzed by using a suitable modification of this finite-temperature theory, leading to a better determination of the three-body parameter in this system~\cite{3BP_li6}.

The Efimov period is significantly smaller for strongly mass-imbalanced heteronuclear systems~\cite{Efimov73}, making them ideal candidates for testing the Efimov discrete scaling invariance~\cite{DIncao2006eto}. Very recently, experiments~\cite{chin_LiCs_efimov,heidelberg_LiCs_efimov} have observed more than one period of the Efimov scaling dependence in the system of one Li and two Cs atoms, where the Efimov period is $\simeq 5$. For $^6$Li-$^{87}$Rb-$^{87}$Rb this quantity is $\simeq 7$ and the Rb-Li mixture is thus also potentially interesting from the viewpoint of testing the Efimov scenario.

In this article we generalize the $S$-matrix formalism developed for three bosons (case BBB) in Refs.~\cite{Efimov1979,RevueBraaten,Braaten_rec_T,UnitaryBoseGas} to the case of an atom interacting with two identical bosons (case ABB). We assume that the AB interaction is tuned to the negative side of a Feshbach resonance and consider two cases for the BB interaction: {\it i} the BB interaction is 
neglected, 
and {\it ii} the BB scattering length is large and negative.
Case {\it i} is suitable
for the Li-Rb-Rb system,
while
case {\it ii} is relevant
for the Li-Cs-Cs one in the magnetic field region studied in Refs.~\cite{chin_LiCs_efimov,heidelberg_LiCs_efimov}. 
In none of these two cases the system supports weakly bound dimers and, therefore, the loss is entirely determined by the recombination to deeply bound states. Under these conditions we express the loss rate constant in terms of the temperature, three-body parameter, inelasticity parameter, and a quantity $s_{11}$ which is a function of a single variable $ka_{AB}$ in case {\it i} and of two variables, $ka_{AB}$ and $ka_{BB}$, in case {\it ii}, where $k$ is the three-body collision momentum. Once this universal function is calculated (and tabulated) for a given AB mass ratio, one can then easily generate loss curves for any given temperature, three-body parameter, and inelasticity parameter.

The article is organized as follows.
In Sec.~\ref{sec:main} we introduce basic notations, present a formula
for the loss rate constant, and apply it to $^6$Li-$^{87}$Rb-$^{87}$Rb
and $^6$Li-$^{133}$Cs-$^{133}$Cs commenting on the role of the BB
interaction. In Sec.~\ref{sec:deriv} we give a detailed derivation of
this loss rate formula.
The calculation of the universal function $s_{11}$ is the subject of Sec.~\ref{sec:s11}. 
Appendices include the normalization constant and contact parameters for Efimov ABB trimers,
the analytic expansion of $s_{11}$ near unitarity, and in App.~\ref{app:homo} we show how one can recover 
the case BBB 
from the present article.

\section{Main results}
\label{sec:main}

We start with some reminders and notations
about
the three-body problem of
two bosons of mass $\mB$ and a third particle of mass $\mA$.
We write
$m$ for twice the reduced mass, 
\be
m = 2 \, \frac{  \mA \mB }{ \mA +\mB}.
\ee
The mass ratio is conveniently parameterized by the angle
\be\label{eq:phi}
\sin \phi = \frac{\mB}{\mA + \mB}.
\ee
For further convenience we also define
\be
\theta = \frac{\pi}{4}-\frac{\phi}{2}. 
\label{eq:def_theta}
\ee
We denote by
$a_{\rm AB}$ and $a_{\rm BB}$
the AB and BB scattering lengths.

Universal properties, such as the Efimov discrete scaling invariance,
appear in the zero-range limit, that can be
described by a universal theory~\cite{Efimov,RevueBraaten2}.
Accordingly,
we consider
zero-range interactions
 between A and B particles,
while between B particles we assume no interactions in case {\it i}
 and zero-range interactions in case {\it ii}.
The validity conditions for this universal zero-range theory are the following.
In case {\it i}, 
the characteristic interaction
ranges and $|a_{\rm BB}|$ should be much smaller than $|a_{\rm AB}|$ and the typical de Broglie wavelengths.
In case {\it ii}, 
the characteristic interaction
ranges should be much smaller than $|a_{\rm AB}|$, $|a_{\rm BB}|$, and the typical de Broglie wavelengths.

The {\it unitary limit} is defined by
\begin{equation}\label{eq:unitary_limit}
\renewcommand\arraystretch{1.5}
\begin{array}{l@{}l}
    a_{\rm AB}= \infty &{}\hspace{1cm} {\rm in\, case}\, i,\\
    a_{\rm AB}=a_{\rm BB} = \infty &{}\hspace{1cm} {\rm in\, case}\, ii.
\end{array}
\end{equation}
As usual, scattering length values $+\infty$ and $-\infty$ are equivalent.
At the unitary limit, 
the scattering length(s) do not introduce any lengthscale into the problem.
Due to the Efimov effect, there is no continuous scale invariance, but a discrete one. In particular,
the energies of Efimov trimers follow a geometric series,
\be
E_{n+1}/E_n = e^{- 2 \pi / s_0}.
\ee
Here 
$s_0$ is the positive real solution of~\cite{Efimov73}
\bea
\lambda(s_0, \phi) &=& 1   \ \ \ \ \ {\rm in\ case}\ i,
\label{eq:s}
\\
\!\!\!\! 1-\lambda(s_0,\phi)-2\lambda^2(s_0,\theta)&=&0
\ \ \ \ \ {\rm in\ case}\ ii,
\label{eq:s_with_BB}
\eea
where
\be
\lambda(s_0,\delta)\equiv \frac{2\,\sinh(\delta s_0)}{s_0\cosh(\pi s_0/2)\sin(2\delta)}.
\label{eq:def_lambda}
\ee

Note that we do not consider here the case of a very large mass ratio, where additional Efimovian sectors would appear in non-zero total angular momentum subspaces. We thus assume that $\mB/\mA$ is smaller than the critical value $\simeq 40$ where an additional Efimov effect appears in the angular momentum $L=2$ subspace~\cite{Efimov73}.

The Efimov effect also implies that the interactions within the universal zero-range theory are parameterized not only by
the scattering lengths, but also by a three-body parameter. For this purpose we use a length parameter $R_0$ defined modulo multiplication by $e^{\pi/s_0}$. This parameter fixes a hyperradial node of the three-body wavefunction at small interparticle distances and thus fixes all other three-body observables [see Eq.~(\ref{eq:Eefi}) for its relation to the spectrum of Efimov trimers]. To account for recombination losses to deeply bound states in the zero-range theory one allows the three-body parameter to be complex, $R_0\rightarrow R_0\,e^{-i \eta / s_0}$, where $\eta>0$ is the inelasticity parameter~\cite{RevueBraaten2}. A precise definition of $R_0$ and $\eta$ is given by the three-body contact condition Eq.~(\ref{eq:3body_cc}) below.

Summarizing, the external parameters of the zero-range theory are the scattering length $a_{\rm AB}$, three-body parameter $R_0$, inelasticity parameter $\eta$, and also the scattering length $a_{\rm BB}$ in case {\it ii}.
Experimentally, all these parameters depend on the magnetic field $B$.
However, essential for universal Efimov physics is only the resonant enhancement of $a_{\rm AB}$ near a Feshbach resonance. In its sufficiently narrow vicinity the other parameters can be assumed constant since their $B$-dependence is smooth. It is important to note that the assumption of constant $R_0$ and $\eta$ is not directly related to the applicability conditions of the zero-range approximation mentioned above. 
In particular, all our derivations and formulas remain valid for $B$-dependent $R_0$ and $\eta$. However, these parameters are assumed constant in Figs.~1 and 2.


\subsection{Loss rate constant}

Our system of interest is
 a mixture of A and B atoms,
in the gaseous non-degenerate regime where the thermal wavelength is small compared to the interparticle distances.
We do not consider finite positive values of the scattering lengths, where weakly bound dimers would exist;
we also assume that $a_{\rm AB}$ is non-zero, in order to have a non-trivial three-body problem.

The rate of ABB recombination events per unit volume is $K\,n_{\rm A}\,n_{\rm B}^{\phantom{B}2}$, where $n_{\rm A}$ and $n_{\rm B}$ are the atom number densities, and $K$ is the event rate constant, for which we obtain
\begin{multline}
K = 64 \, \pi^2 \cos^3 \! \phi \,\, \frac{\hbar^7}{m^4 (k_B T)^3}
\left( 1 - e^{-4 \eta} \right)
\\
\times
\int_0^\infty
\frac{1 - |s_{11}|^2}{|1 + (k R_0)^{-2 i s_0} e^{-2\eta} \, s_{11} |^2}
\,e^{-\hbar^2k^2/mk_BT}
\,
k \, dk.
\label{eq:K_T}
\end{multline}
The function $s_{11}$ depends only on $k a_{\rm AB}$, $k a_{\rm BB}$, and the mass ratio. 
Therefore, Eq.~(\ref{eq:K_T}) gives an explicit dependence of the loss rate on the three-body and inelasticity parameters.
This is a manifestation of a general property known as Efimov's radial law~\cite{Efimov1979,RevueBraaten}. 

For high temperatures or large scattering lengths such that in case {\it i} we typically have $-k a_{\rm AB}\gg 1$ and in case {\it ii} $-k a_{\rm AB}\gg 1$ and $-k a_{\rm BB}\gg 1$, the function $s_{11}$ in Eq.~(\ref{eq:K_T}) can be approximated by its asymptotic unitary value
\be
s_{11}^\infty = - e^{ -\pi s_0 + i\, 2 \left[ s_0 \ln 2 + {\rm arg} \, \Gamma( 1 + i s_0 ) \right]}
\label{eq:s11_unitary}
\ee 
with $s_0$ given by Eqs.~(\ref{eq:s},\ref{eq:s_with_BB}). 

In Section~\ref{sec:s11} we show how to compute $s_{11}$ at finite $k a_{\rm AB}$ and $k a_{\rm BB}$ for any given mixture. Once this task is accomplished, Eq.~(\ref{eq:K_T}) offers a very fast way of calculating the loss rate for any $T$, $R_0$, and $\eta$, and can thus be used for extracting these parameters from experimental data in the universal limit. Obviously, in this manner not only experimental but also theoretical results obtained for finite-range potentials can be compared to the zero-range theory.

In the next two subsections we employ Eq.~(\ref{eq:K_T}) for calculating $K$ in two experimentally relevant cases characterized by large mass ratios: ${\rm A=\,^6Li}$, ${\rm B}=\,^{87}{\rm Rb}$ and ${\rm A=\,^6Li}$, ${\rm B}=\,^{133}{\rm Cs}$.~\footnote{Tabulated values of $s_{11}$ for these systems can be obtained by contacting the authors.}

\subsection{Lithium-Rubidium}

We first consider the case ${\rm A}=\,^6$Li, ${\rm B}=\,^{87}{\rm Rb}$ and neglect the $^{87}$Rb-$^{87}$Rb interaction since it is non-resonant apart from narrow magnetic field intervals~\cite{Marte2002fri,FeshbachRMP2010}. Broad interspecies Feshbach resonances in the hyperfine ground states are available for both $^6{\rm Li}$-$^{87}{\rm Rb}$ and $^7{\rm Li}$-$^{87}{\rm Rb}$ combinations \cite{ZimmermannLi6RbRes,ZimmermannLi7RbRes,FeshbachRMP2010}. However, for the purpose of testing the Efimov scenario in the ABB system the choice of fermionic A allows one to neglect concomitant AAB losses suppressed due to the fermionic statistics~\cite{PetrovJPhysB}.

\begin{figure}
\centerline{\includegraphics[width=\hsize,clip,angle=0]{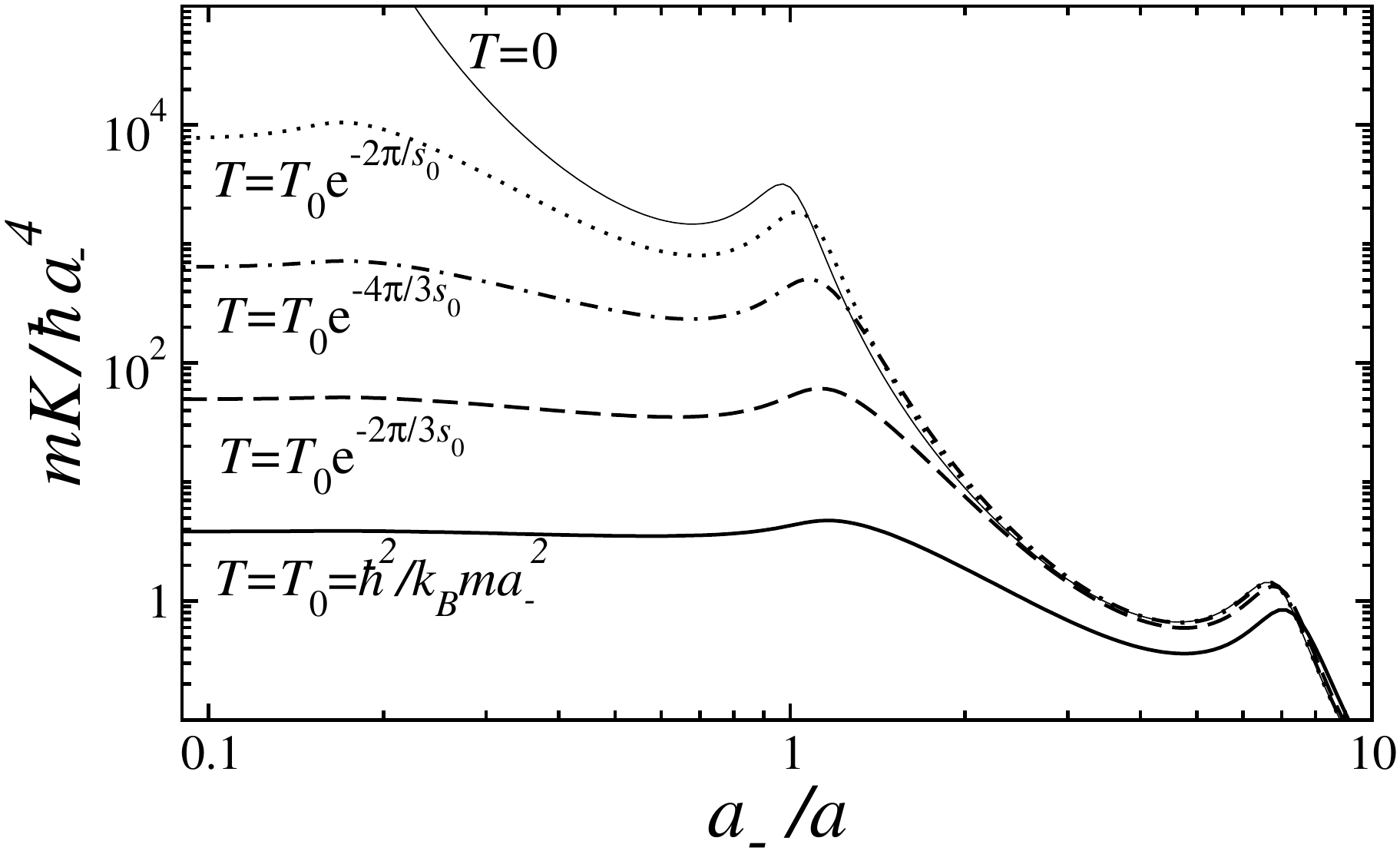}}
\caption{The event rate constant $K$ in units of $\hbar a_{-}^4/m$ versus $a_{-}/a$ for the $^6{\rm Li}$-$^{87}{\rm Rb}$-$^{87}{\rm Rb}$ system for various temperatures with the inelasticity parameter set to $\eta=0.2$.}
\label{fig:LiRbRb}
\end{figure}

The $^6{\rm Li}$-$^{87}{\rm Rb}$-$^{87}{\rm Rb}$ system is characterized by $s_0 = 1.63188$ and a scaling factor $e^{\pi/s_0}=6.85610$. In Fig.~\ref{fig:LiRbRb} we present the event rate constant $K$ in units of $\hbar a_{-}^4/m$ as a function of $a_{-}/a$ for different temperatures. Here $a$ is the Rb-Li scattering length and $a_{-}<0$ is the position of the three-body loss peak at zero temperature (at $a=a_{-}$ an Efimov trimer crosses the three-atom threshold). The (three-body) parameter $a_{-}$ is convenient when there is only one relevant scattering length (for instance, for three identical bosons or for our case {\it i}). We find that for the $^6{\rm Li}$-$^{87}{\rm Rb}$ mass ratio $a_{-}$ is related to $R_0$ by $a_{-}\approx -5.233 R_0$. 

The shapes of the curves are controlled by $\eta$ and $T$. Unaware of experimental results, we set $\eta=0.2$. The lowest curve in Fig.~\ref{fig:LiRbRb} corresponds to $T=T_0=\hbar^2/k_Bma_{-}^2$ and the other ones are obtained by decreasing $\ln T$ by $2\pi/3s_0$, i.e., a third of the Efimov energetic period. Thus, the solid curve for $T=T_0$ and the dotted one for $T=T_0e^{-2\pi/s_0}$ are self-similar, the latter can be obtained by shifting the former horizontally by $e^{-\pi/s_0}$ and vertically by $e^{4\pi/s_0}$. The zero-temperature result (thin solid) is known analytically \cite{HelfrichHammerPetrov}. The shift of the resonance peak towards smaller $|a|$ with increasing $T$ is likely to be related to the fact that for $|a|<|a_{-}|$ the trimer becomes a finite-energy resonance in the three-atom continuum.

\subsection{Lithium-Cesium}

Because of its larger mass ratio, the combination A=$^6$Li, B=$^{133}$Cs is considered as an even better candidate for observing several periods of the Efimov discrete scaling. Two groups have recently reported loss measurements for this mixture close to a wide CsLi Feshbach resonance at 843G \cite{chin_LiCs_efimov,heidelberg_LiCs_efimov}. The first two Efimov resonances and signatures of the third one have been detected, although the latter is rather strongly thermally saturated.

In treating the $^6$Li-$^{133}$Cs-$^{133}$Cs system for magnetic fields studied in \cite{chin_LiCs_efimov,heidelberg_LiCs_efimov} one has to have in mind that the CsCs scattering length is rather large (about -1550$a_{\rm Bohr}$ at the CsLi resonance \cite{Berninger2013frw}) and this may increase the loss rate constant because of the enhanced probability to find two Cs atoms close to each other. On the other hand, to describe the Efimov effect for such a large mass ratio one is tempted to use the Born-Oppenheimer approximation~\cite{FonsecaBO1979}, within which the heavy-heavy interaction is irrelevant for the discrete scaling. Indeed, the $^6$Li-$^{133}$Cs-$^{133}$Cs system in case {\it i} is characterized by a scaling factor $e^{\pi/s_0}= 4.87661$ [Eq.~(\ref{eq:s}) gives $s_0= 1.98277$], whereas for  $a_{\rm CsCs}=\infty$ this quantity becomes 4.79887 [$s_0= 2.00308$ as follows from Eq.~(\ref{eq:s_with_BB})] (cf. \cite{Yamashita2013spm}).  

\begin{figure}
\centerline{\includegraphics[width=\columnwidth]{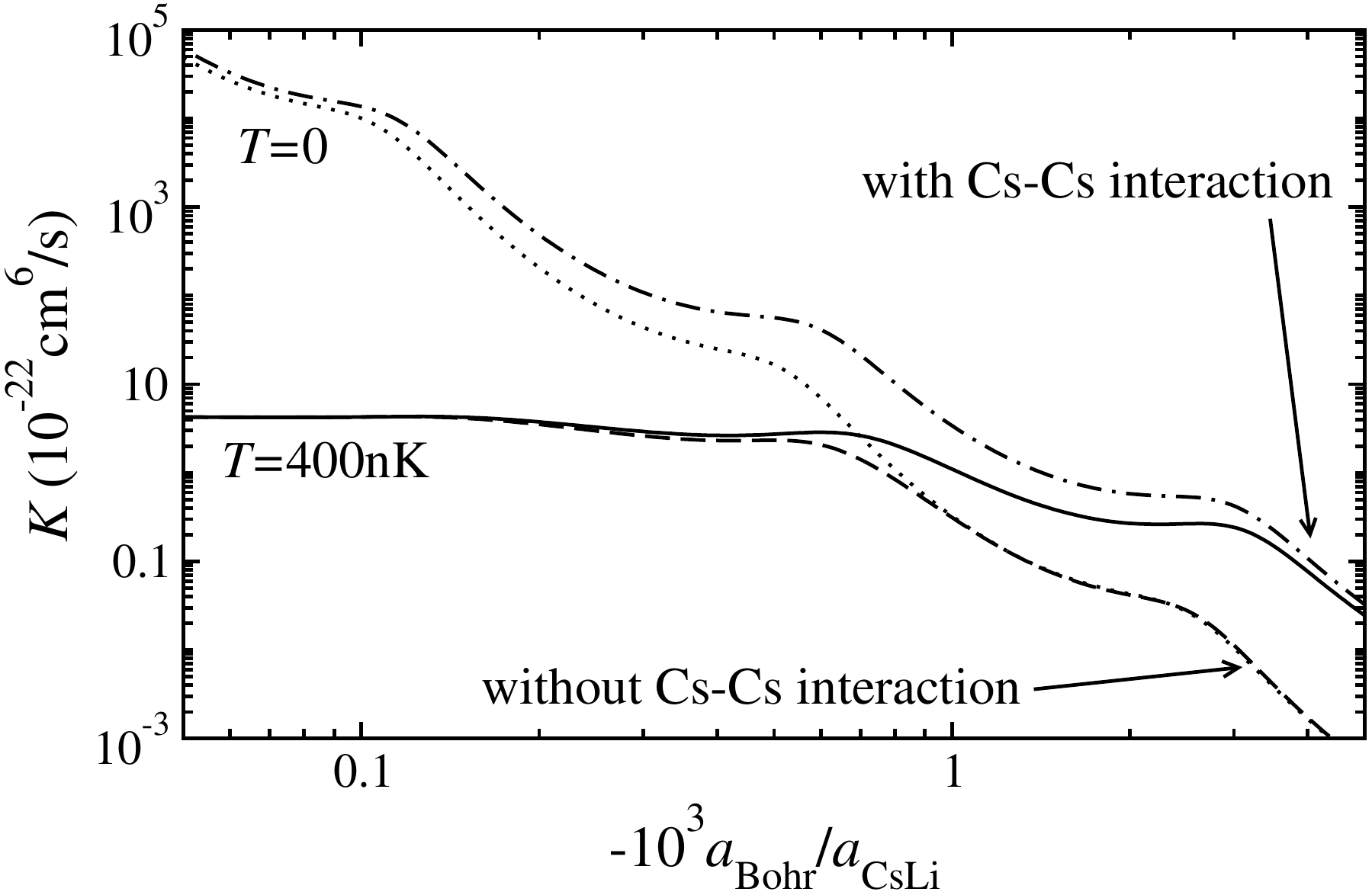}}
\caption{
$K$ vs. $-1/a_{\rm CsLi}$ at $T=0$ and $T=400$nK including and excluding the CsCs interaction. The three-body and inelasticity parameters are set to 
 $R_0=130a_{\rm Bohr}$
 and $\eta=0.6$, respectively.
} 
\label{fig:KLiCsCs}
\end{figure}

In Fig.~\ref{fig:KLiCsCs} we plot $K$ vs $-1/a_{\rm CsLi}$ for $T=400$nK (solid and dashed) and $T=0$ (dash-dotted and dotted) with 
 $R_0=130a_{\rm Bohr}$ 
and $\eta=0.6$. Including the CsCs interaction we obtain the solid and dash-dotted curves whereas neglecting it gives the dashed and dotted lines. In the CsCs-interacting case we use the magnetic field dependence $a_{\rm CsCs}(B)$ provided in Suplemental Material of Ref.~\cite{Berninger2013frw} and eliminate the magnetic field by using the formula $a_{\rm CsLi}=-28.5a_{\rm Bohr}[61.4{\rm G}/(B - 842.9{\rm G}) + 1]$ \cite{heidelberg_LiCs_efimov}. A different $a_{\rm CsLi}(B)$ was used in Ref.~\cite{chin_LiCs_efimov} and Ref.~\cite{heidelberg_LiCs_Fesh_Res} gives yet another more recent characterization of $a_{\rm CsLi}(B)$. For the results plotted in Fig.~\ref{fig:KLiCsCs} these variations are not very important because $a_{\rm CsCs}(B)$ is smooth and, therefore, the dependence $a_{\rm CsCs}(a_{\rm CsLi})$ is practically unchanged.

Figure~\ref{fig:KLiCsCs} clearly shows that the effect of finite $T$ is to saturate $K$ at large $a_{\rm CsLi}$ whereas the inclusion of the CsCs interaction leads to a strong enhancement of the loss rate for small $a_{\rm CsLi}$. For the experimentally studied region of $B$ either of these effects can lead to corrections of two orders of magnitude. The overall behavior of $K$ is thus much less steep than the scaling $K\propto a_{\rm CsLi}^4$ (compare solid and dotted lines in Fig.~\ref{fig:KLiCsCs}), consistent with the experimental findings~\cite{chin_LiCs_efimov,heidelberg_LiCs_efimov}. The peak positions move when we ``switch on'' the CsCs interaction, but this is an artefact of choosing the same $R_0$ in these two cases. In fact, even in the limit $a_{\rm CsLi}\rightarrow \infty$ the loss features are not expected to exactly match. This is because the Efimovian type-{\it ii} wave function has to propagate through distances of order $a_{\rm CsCs}$ before it can be matched with the type-{\it i} one.  In general, when both cases {\it i} and {\it ii} are applicable, i.e., when $|a_{\rm BB}|$ is much larger than the characteristic interaction ranges but smaller than $|a_{\rm AB}|$ and $1/k$, the corresponding three-body parameters are related to each other by the wave-function matching condition, but are not necessarily equal.

The parameters 
$R_0=130a_{\rm Bohr}$
 and $\eta=0.6$ used for plotting Fig.~\ref{fig:KLiCsCs} have been chosen as a result of a very approximate fitting of the data of Refs.~\cite{heidelberg_LiCs_efimov,chin_LiCs_efimov}. A significantly more serious account of experimental uncertainties and cross-correlations is needed to give a more definite answer for $R_0$ and $\eta$. We find that the fitting procedure is very sensitive to the exact position of the Feshbach resonance. Surely, one would benefit from more experimental data at lower temperatures. Setting $T=0$, $\eta\ll 1$, and 
$R_0=130a_{\rm Bohr}$
 we obtain sharp peaks of $K$ at positions  $a_{\rm CsLi}=a_{-}^{(0)}\approx -330a_{\rm Bohr}$, $a_{\rm CsLi}=a_{-}^{(1)}\approx -1.7\times 10^3a_{\rm Bohr}$, $a_{\rm CsLi}=a_{-}^{(2)}\approx -8.8\times 10^3a_{\rm Bohr}$, which compares very well with the refined analysis of the experimental data performed in Ref.~\cite{heidelberg_LiCs_Fesh_Res}. 

We note that the experiments~\cite{heidelberg_LiCs_efimov,chin_LiCs_efimov} are well within the zero-range limit, at least, for the second and higher Efimov resonances. Indeed, out of the van der Waals ranges \cite{FeshbachRMP2010,c6_hetero}, the Feshbach resonance parameters $R_*$ \cite{PetrovLesHouches2010,FeshbachRMP2010,chin_LiCs_Feshb_Res_2013}, and the length $\sqrt{ R_* |a_{\rm bg}|}$ that enters when the background scattering length $a_{\rm bg}$ is negative~\cite{LeChapitre}, the largest is the Cs-Cs van der Waals range $r_{\rm vdW,CsCs}=101a_{\rm Bohr}$. In turn, this quantity is much smaller than the thermal wavelengths (${>}3400\,a_{\rm Bohr}$ at $T<800\,{\rm nK}$) and the Cs-Cs scattering length, which, for experimentally relevant magnetic fields, varies in the interval $-1550a_{\rm Bohr}<a_{\rm CsCs}<-900a_{\rm Bohr}$ \cite{Berninger2013frw}. As far as the Cs-Li scattering length is concerned, around the first Efimov resonance it approximately equals $-300a_{\rm Bohr}$, which is comparable to $r_{\rm vdW,CsCs}$. However, for the second and higher resonances the inequality $|a_{\rm CsLi}|\gg r_{\rm vdW,CsCs}$ is well satisfied.

Wang {\it et al.}~\cite{Wang2012utb} have studied the so-called van der Waals universality of the three-body parameter in heteronuclear systems close to a wide interspecies resonance by assuming the Lennard-Jones interatomic potentials. According to their analytical estimates in the case of a large mass imbalance the three-body parameter equals the heavy-heavy van der Waals range $r_{\rm vdW,BB}$ times a dimensionless function of the ratio $r_{\rm vdW,BB}/a_{\rm BB}$. For the Cs-Cs-Li case in the experimentally relevant region this function varies very little ($<2$\%) since the ratio $r_{\rm vdW,CsCs}/a_{\rm CsCs}$ stays small. We mention for reference that for $a_{\rm CsCs}=2000a_{\rm Bohr}$ Ref.~\cite{Wang2012utb} predicts $a_{-}=-1.4\times 10^3a_{\rm Bohr}$.

A finite-temperature theoretical analysis of losses in this system has been performed by Y.~Wang and reported in the Supplemental Material of Ref.~\cite{chin_LiCs_efimov}. Wang's results indicate that with increasing temperature the resonance features become weaker and shift towards smaller $a_{\rm CsLi}$, which is what we also observe. In fact, we can rather well fit Wang's curves for all three considered temperatures ($T=100$nK, 250nK, and 1$\mu$K) by choosing 
$R_0=110a_{\rm Bohr}$
and $\eta=0.4$.

\section{Derivation of the loss rate constant}
\label{sec:deriv}

In this section, we derive the expression (\ref{eq:K_T}) for the loss rate constant in terms of the quantity $s_{11}$, as well as the analytical result~(\ref{eq:s11_unitary}) for $s_{11}$ valid at the unitary limit. 

Let us first introduce
 the three-body scattering state $\psi$.
Denoting the positions of the two identical bosons by $\rr_1$ and $\rr_3$, and the position of the third particle by $\rr_2$,
the wavefunction $\psi(\rr_1,\rr_2,\rr_3)$ is symmetric with respect to exchanging $\rr_1$ and $\rr_3$.

Let us introduce the
Jacobi coordinates 
\begin{equation}\label{eq:def_xy}
\renewcommand\arraystretch{1.8}
\begin{array}{r@{}l}
    \cos \phi\ \xx  &{}= \rr_3 - \frac{\mA \rr_2 + \mB \rr_1}{\mA+\mB},\\
    \yy&{} =  \rr_2 - \rr_1.    
\end{array}
\end{equation}
All information about the relative positions of the three particles can then be collected in the 6-dimensional vector
\be
\RR = (\xx,\yy).
\ee
Its norm $R=\sqrt{x^2+y^2}$ is the so-called hyperradius, while its direction 
$$\Oo \equiv \RR/R$$
 can be parameterized by five hyperangles.
One can note that 
if all particle coordinates are multiplied by a factor $\lambda$, then
the hyperradius is multiplied by $\lambda$ while the hyperangles are unchanged.

At large distances, the scattering state asymptotes to an incoming plane wave plus a scattered wave. More precisely, in the center-of-mass reference frame,
\be
\psi(\RR) \underset{R\to\infty}{\simeq} \psi^{(0)}(\RR) + \psi_{\rm sc}(\RR)
\label{eq:psi_psi0}
\ee
where
 $\psi_{\rm sc}(\RR)$ is a purely outgoing scattered wave,
and 
\be
\psi^{(0)}(\RR)=\frac{1+P}{\sqrt{2}}\, e^{i \kk \cdot \RR} 
\label{eq:sym_eikR}
\ee
is a symmetrized plane wave, normalized in a unit volume. The operator $P$ in Eq.~(\ref{eq:sym_eikR}) exchanges particles $1$ and $3$.
The relation between $k$ and the collision energy is
\be
E=\frac{k^2}{m}.
\ee
Here and in what follows, we set $\hbar=1$.

Furthermore, $\psi$ is
an eigenstate of the
zero-range model, 
i.e., it satisfies
\bi
\item
the Schr\"odinger equation
\be
-\frac{1}{m} \Delta_{\RR} \psi(\RR) = E \, \psi(\RR)
\label{eq:schro}
\ee
when none of the three particle positions coincide;
\item
the two-body contact conditions: for each interacting pair of particles $i, j$,
$\exists A_{ij}$ such that
\be
\psi \underset{r_{ij}\to0}{=} \left( \frac{1}{r_{ij}} - \frac{1}{a_{ij}} \right) \, A_{\ij} + o(1)
\label{eq:2body_cc}
\ee
where $A_{\ij}$ depends on the relative position of the third particle with respect to  the center-of-mass of particles $i$ and $j$;
\item
the three-body contact condition: $\exists B$ such that
\be
\psi(\RR) \underset{R\to0}{\sim}
\left[
\left(\frac{R}{R_0}\right)^{-i s_0}
-
e^{-2\eta}
\left(\frac{R}{R_0}\right)^{i s_0}
\right]
\,
\frac{B(\Oo)}{R^2}.
\label{eq:3body_cc}
\ee
\ei

The loss rate can then be obtained from the scattering state thanks to the following exact relation,
that can be justified by heuristic arguments.
Introducing the probability current
\be
\JJ = \frac{2}{m} \, {\rm Im} \left( \psi^* \, \gr_{\RR} \psi \right)
\label{eq:def_J}
\ee
and the lost flux
\be
\varphi_{\rm loss} = - \oint_{\Sr} \JJ \cdot \mathbf{d^5 S}
\label{eq:def_phi_loss}
\ee
where $\Sr$ is a hypersurface enclosing the origin (e.g. a hypersphere)
and 
the surface-element vector
$\mathbf{d^5 S}$ points away from the origin,
the loss rate constant is given by the thermal average
\be
K = \frac{\int K(k) \, e^{-k^2/mk_BT} \, d^6k}{\int  e^{-k^2/mk_BT} \, d^6k}
\ee
of the energy-resolved event rate constant
\be
K(k) = \frac{\cos^3 \! \phi}{2} \, \langle \varphi_{\rm loss} \rangle_{\hat{k}}
\label{eq:K_k_vs_loss}
\ee
where $\langle . \rangle_{\hat{k}}$ denotes the average over the direction of $\kk$.
In Eq.~(\ref{eq:K_k_vs_loss}),
the factor $1/2$ originates from the indistinguishability of the two B particles, and
the factor $\cos^3 \! \phi$   from the Jacobian of the change of variables from Cartesian to Jacobi coordinates,
\be
\left|
\frac{\partial(\rr_1,\rr_2,\rr_3)}{\partial(\CC,\xx,\yy)}
\right| = \cos^3\!\phi
\label{eq:jacobian}
\ee
where $\CC$ is the center-of-mass of the three particles.

\subsection{Unitary limit} \label{sec:unitary}

Let us first consider the unitary limit
Eq.~(\ref{eq:unitary_limit}),
where the situation is particularly clear, because the problem can be solved in a fully analytical way.
The key ingredient is that there is a separability between the hyperradius $R$ and the hyperangles $\Oo$,
because the two-body contact conditions~(\ref{eq:2body_cc}) do not introduce any lengthscale and hence act only on the hyperangles~\cite{Efimov73}.
The solutions of the hyperangular part of the three-body problem are the functions $\phi_s(\Oo)$ that satisfy the two-body contact condition and which are eigenfunctions of the Laplacian operator on the hypersphere:
\be
T_\Oo \, \phi_s(\Oo) = -s^2 \phi_s(\Oo).
\label{eq:eigen_hyperang}
\ee
The operator $T_\Oo$ is defined as the hyperangular part of the total Laplacian
\be
\Delta_{\RR} = \frac{1}{R^2} \left(\frac{\partial^2}{\partial R^2} + \frac{1}{R} \frac{\partial}{\partial R} + \frac{1}{R^2} T_{\Oo} \right) R^2.
\label{eq:T_Oo}
\ee
 We can then expand the scattering state as
\be
\psi(\RR) = \sum_s \frac{F_s(R)}{R^2} \, \phi_s(\Oo).
\label{eq:sum_s}
\ee
Indeed, the functions $\phi_s$,
normalized to unity,
form an orthogonal basis
for the hyperangular scalar product
\be
(f|g) \equiv \int f(\Oo)^* \, g(\Oo) \, d\Oo
\label{eq:def_ps}
\ee
(where $d\Oo$ stands for the differential solid angle in six-dimenional space, i.e. $d^6 R = d\Oo \ R^5 \, dR$). This follows from the self-adjointness of the zero-range model.

In the present case,
where the B particles are bosonic and the mass ratio is not very large,
the set $\{s\}$ contains a single imaginary value $s = i s_0$ where $s_0$ solves Eq.~(\ref{eq:s}), and an infinite countable set of real values.
The $s = i s_0$ sector is called Efimovian,
since it causes the Efimov effect.

The hyperradial Schr\"odinger equation reads
\be
\left( -\frac{d^2}{dR^2} - \frac{1}{R} \frac{d}{dR} + \frac{s^2}{R^2} \right) F_s(R) = 
m E\,
F_s(R),
\ee
i.e., the unitary three-body problem reduces to a set of independent one-body problems in effective $s^2/R^2$ potentials.
The hyperradial wavefunctions $F_s(R)$
have the large-distance behavior
\be
F_s(R) \underset{R\to\infty}{\simeq} \left[
A^{\rm in}(s)\, e^{- i k R}
+
A^{\rm out}(s) \, e^{i k R}
\,
\right]
\sqrt{\frac{m}{2 k R}},
\label{eq:in_and_out_amps}
\ee
where a normalisation to unit flux is introduced for later convenience.
For the real values of $s$, due to the repulsive effective potential $s^2/R^2$, the wavefunction $F_s(R)$ vanishes for $R\to0$
(in the absence of three-body resonance),
and $|A^{\rm in}(s)|=|A^{\rm out}(s)|$, i.e., the scattering is purely elastic.
The losses thus come exclusively from the Efimovian sector.
There, the strongly attractive effective potential $-s_0^2/R^2$ gives rise to  logarithmic waves at small hyperradius (i.e., in the limit where all three particles are close):
\be
F_{i s_0}(R) \underset{R\to0}{\simeq} \left[ A_1^{\rm in} \, e^{i s_0 \ln(kR)}
+
A_1^{\rm out} \, e^{-i s_0 \ln(kR)}
\right]
\,
\sqrt{\frac{m}{2 s_0}},
\ee
where a unit-flux normalisation is again introduced for later convenience.
The three-body contact condition Eq.~(\ref{eq:3body_cc}) becomes
\be
A_1^{\rm in} = \Ar \, A_1^{\rm out},
\label{eq:Ar}
\ee
where
\be
\Ar \equiv - (k R_0)^{-i 2 s_0} \, e^{-2 \eta}
\label{eq:Ar=}
\ee
has the meaning of a reflection amplitude from the point $R=0$ (where all particle positions coincide).
While the phase of this reflection amplitude is determined by the three-body parameter $R_0$,
its modulus is determined by the inelasticity parameter $\eta$,
the reflection probability being $|\Ar|^2 = e^{-4\eta}$.
Accordingly,
\be
\varphi_{\rm loss} = (1 - e^{-4\eta}) \, |A^{\rm out}_1|^2
\label{eq:phi_loss_vs_A1out}
\ee
[as obtained by taking a vanishingly small hypersphere for $\Sr$ in Eq.~(\ref{eq:def_phi_loss})].

In what follows we denote by
$A_3^{\rm in/out}$ the long-distance amplitudes $A^{\rm in/out}(i s_0)$ [see Eq.~(\ref{eq:in_and_out_amps})].~\footnote{
The subscript $3$ is used here since the subscript $2$ is traditionally reserved for the atom weakly-bound-dimer channel~\cite{Efimov1979,RevueBraaten,Braaten_rec_T}.
}
The out-amplitudes can be expressed in terms of the
in-amplitudes through a linear relation,
\be
\begin{pmatrix}
A_1^{\rm out}
\\
A_3^{\rm out}
\end{pmatrix}
=
\begin{pmatrix}
s_{11} & s_{13}
\\
s_{31} & s_{33}
\end{pmatrix}
\begin{pmatrix}
A_1^{\rm in}
\\
A_3^{\rm in}
\end{pmatrix}.
\label{eq:2by2}
\ee
Combining this with Eq.~(\ref{eq:Ar}) yields
\be
A_1^{\rm out} = \frac{ s_{13}}{1 - s_{11} \Ar} \,\, A^{\rm in}_3.
\label{eq:A1out_vs_A3in}
\ee
The $S$-matrix $s_{ij}$ is easily computed by using the fact that $F_{i s_0}(R)$ is a linear combination of the Bessel functions $J_{\pm i s_0}(kR)$.
This yields the expression of $s_{11}$ given in
 Eq.~(\ref{eq:s11_unitary}).
Furthermore the
 obtained matrix 
 is unitary, so that $|s_{13}|^2 = 1 - |s_{11}|^2$ and thus
\be
\varphi_{\rm loss} = \rP |A^{\rm in}_3|^2,
\label{eq:loss_vs_A3in}
\ee
where the recombination loss probability $\rP$ equals (compare with Ref.~\cite{WangNJP2011})
\begin{equation}\label{eq:loss_probability}
\rP= (1 - e^{-4\eta})\, \frac{ 1-|s_{11}|^2}{|1 - s_{11} \Ar|^2}.
\end{equation}

The last ingredient is that the large-distance in-amplitude $A^{\rm in}_3$
is determined by the projection of the incoming plane-wave onto the Efimovian sector.
More precisely,
the hyperangular overlap 
$(\phi_{i s_0}|\psi^{(0)}) = \sqrt{2} \, \int d^5 \Oo \, \phi_{i s_0}(\Oo) \, e^{i \kk \cdot \RR} $
can be evaluated for $R\to\infty$ using the stationary-phase method,
with a result of the form 
\be
[ A^{{\rm in}\, (0)}_3 \, e^{-i k R} + A^{{\rm out}\, (0)}_3 \, e^{i k R} ]
\sqrt{\frac{m}{2 k }} \, R^{-5/2};
\label{eq:form}
\ee
then, $A^{{\rm in}}_3 - A^{{\rm in}\, (0)}_3$ must vanish,
because
$(\phi_s | \psi - \psi^{(0)})$ has to behave like a purely outgoing wave at $R\to\infty$, by definition of the scattering state $\psi$ [cf. Eq.~(\ref{eq:psi_psi0})].
This yields
\be
A_3^{\rm in} =  2^{7/2} \, \pi^{5/2} \, e^{i\, 5\pi/4} \, m^{-1/2} \, k^{-2} \,  \phi_{i s_0}\!(-\hat{k})^* .
\label{eq:A3in}
\ee

Inserting this into Eqs.~(\ref{eq:loss_vs_A3in}) and (\ref{eq:K_k_vs_loss}),
$\phi_s$ drops out  after the hyperangular average:
\be
K(k) = 
\frac{64\,\pi^2\,\cos^3\!\phi}{m k^4}
\,
(1-e^{-4\eta})
\,
\frac{1-|s_{11}|^2}{|1-s_{11}\Ar|^2}.
\label{eq:K_k}
\ee
The final expression Eq.~(\ref{eq:K_T})  follows immediately.

\subsection{Finite scattering lengths} \label{sec:finite_a}

We
turn to the general case where
the scattering length(s) are not restricted to the unitary limit Eq.~(\ref{eq:unitary_limit}). 
We will see that the final expression for the loss rate Eq.~(\ref{eq:K_T}) remains valid, provided the definition of $s_{11}$ is appropriately generalized.
Since the scale invariance of the two-body contact conditions is broken,
the separability in hyperspherical coordinates no longer holds.
Instead, we follow an $S$-matrix approach.

We consider the state $\Psi_1$ that physically corresponds to a stationary triatomic flow injected at the origin of the six-dimensional space (i.e., at $\RR=\vn$) that gets partially reflected back and partially  transmitted towards infinity.
More precisely,
$\Psi_1$ is defined as the solution of the Schr\"odinger equation~(\ref{eq:schro}) with energy $E=k^2/m$ satisfying the two-body contact condition~(\ref{eq:2body_cc})
and having the asymptotes
\begin{equation}
\Psi_1(\RR)  \underset{R\to0}{\simeq} 
\frac{
(kR)^{i s_0} + s_{11} (kR)^{-i s_0}}{R^2}\sqrt{\frac{m}{2 s_0}}\,\phi_{i s_0}(\Oo)
\label{eq:Psi1_smallR}
\end{equation}
and
\begin{equation}
\Psi_1(\RR)  \underset{R\to\infty}{\simeq}  s_{31} \, e^{i k R} \, \sqrt{\frac{m}{2 k R}} \, \frac{1}{R^2} \, \Phi_3(\Oo).
\label{eq:Psi1_largeR}
\end{equation}
Together with the normalization $(\Phi_3|\Phi_3)=1$,
this defines 
the reflection and transmission amplitudes
$s_{11}$ and $s_{31}$, 
as well as the function $\Phi_3(\Oo)$ 
(up to multiplication of $s_{31}$ and $\Phi_3$ by arbitrary phase factors $e^{i\gamma}$
and
  $e^{-i\gamma}$, respectively).
Equation~(\ref{eq:Psi1_smallR}) is applicable
in the {\it scale-invariant region}, which we can define by $R\ll {\rm min}(|a_{\rm AB}|,1/k)$ in case {\it i} and $R\ll {\rm min}(|a_{\rm AB}|,|a_{\rm BB}|,1/k)$ in case {\it ii}. 
Equation~(\ref{eq:Psi1_largeR}) is applicable
in
the {\it asymptotic region}: $R\gg {\rm min}(|a_{\rm AB}|,1/k)$ in case {\it i} and $R\gg {\rm min}[{\rm max}(|a_{\rm AB}|,|a_{\rm BB}|),1/k]$ in case {\it ii}.

Then, let $\left\{ \Phi_n(\Oo) \right\}_{n\geq4}$  be an arbitrary orthonormal set of functions such that $\left\{ \Phi_n(\Oo) \right\}_{n\geq3}$  forms an orthonormal basis [for the scalar product $(.|.)$ defined in Eq.~(\ref{eq:def_ps})].
A complete set of incoming and outgoing asymptotic states
can be defined as
\bea
\psi_1^{\rm in}  & \equiv &  e^{i s_0 \ln(k R)} \, \sqrt{\frac{m}{2 s_0}}\, \frac{1}{R^2} \,\, \phi_{i s_0}(\Oo)
\\
\psi_n^{\rm in} & \equiv& e^{- i k R}
\, \sqrt{\frac{m}{2 k R}} \, \frac{1}{R^2} \, \Phi_n(\Oo)^*,
\ \ \ n\geq3,
\eea
and $\psi_n^{\rm out} \equiv (\psi_n^{\rm in})^*$ for any $n$ in the set $\Cr \equiv 
\{1\} \cup \{n;\ n\geq3\}$. 
The terms incoming and outgoing are meant with respect to the {\it intermediate region} contained in between the scale-invariant and asymptotic ones. 

For an arbitrary solution $\varPsi$ of Eqs.~(\ref{eq:schro}) and (\ref{eq:2body_cc}),
the in- and out-amplitudes $A_n^{\rm in/out}$ can be defined by
\begin{numcases}{}
     \varPsi    \underset{R\to0}{\simeq} A_1^{\rm in}\,\psi_1^{\rm in} + A_1^{\rm out}\,\psi_1^{\rm out} 
     \label{eq:psi_ampl_0}
     \\
      \varPsi   \underset{R\to\infty}{\simeq}   \sum_{n\geq3
} \left[ A_n^{\rm in}\,\psi_n^{\rm in} + A_n^{\rm out}\,\psi_n^{\rm out} \right].
         \label{eq:psi_ampl}
 \end{numcases}
The out- and in-amplitudes are linearly related:
 \be
 A^{\rm out}_n = \sum_{m\in\Cr}s_{nm} A^{\rm in}_m
 \label{eq:linS}
 \ee
where the matrix $s_{nm}$
is unitary and symmetric, 
as shown in Appendix~\ref{app:s_matrix}.
Furthermore, $s_{nm}$ is independent of the three-body and inelasticity parameters $R_0$ and $\eta$,
and only depends on
 $k a_{\rm AB}$, $k a_{\rm BB}$, and the mass ratio;
indeed,
 we did not impose that $\varPsi$ satisfies Eq.~(\ref{eq:3body_cc}) as the three-body boundary condition.

This problem of an {\it a priori} infinite number of coupled channels actually reduces to only two channels.
Indeed,
the channels $1$ and $3$ decouple from the other ones, i.e. Eq.~(\ref{eq:2by2}) remains valid.
To check this, first note that 
$s_{n1}=0$ for $n\geq 4$ by
construction, cf.~Eq.~(\ref{eq:Psi1_largeR}).
Furthermore, $s_{n3}$ also vanishes for $n\geq4$, because the state
with a purely incoming wave in channel $3$
(denoted by $\Psi_3$ in App.~\ref{app:s_matrix})
is a linear combination of $\Psi_1$ and $\Psi_1^*$.

The rest of the reasoning closely follows the $a=\infty$ case.
Equations~(\ref{eq:Ar},\ref{eq:Ar=},\ref{eq:phi_loss_vs_A1out})
hold, and hence also Eq.~(\ref{eq:A1out_vs_A3in}).
The $s$ matrix being unitary,
Eq.~(\ref{eq:loss_vs_A3in}) follows.
It remains to relate $A_3^{\rm in}$ to the projection of the incoming plane wave onto channel $3$.
For $R\to\infty$,
 the overlap $(\Phi_3^*|\psi^{(0)})$ can again be evaluated
using the stationary-phase method,
with a result of the form
Eq.~(\ref{eq:form}).
On the other hand,
the overlap
$(\Phi_3^*|\psi)$ behaves at large $R$ as
$A_3^{\rm in} \, e^{- i k R} \, \sqrt{m / (2 k R)} \, R^{-2}$
plus an outgoing wave.
Since $(\Phi_3^*|\psi-\psi^{(0)})$ still has a purely outgoing behavior at large $R$, $A_3^{\rm in} - A_3^{\rm in\, (0)}=0$, which finally gives Eq.~(\ref{eq:A3in}) with $\phi_{i s_0}^*$ replaced by $\Phi_3$.
After hyperangular averaging, $\Phi_3$ drops out of the final expressions Eqs.~(\ref{eq:K_k}) and (\ref{eq:K_T}).
This happens because we consider a non-degenerate gas at equilibrium, whose momentum distribution follows the Boltzman law,
so that the three-body momentum distribution depends only on the center-of-mass momentum and on $k$ but not on $\hat{k}$; in general the functional form of $\Phi_3(\Oo)$ does play a role,
see, for example, the study of non-equilibrium effects in Ref.~\cite{chevy_nk_with_losses}.

\subsection{Analogy with an interferometer} \label{sec:fabry}

\begin{figure}
\centerline{\includegraphics[width=\columnwidth]{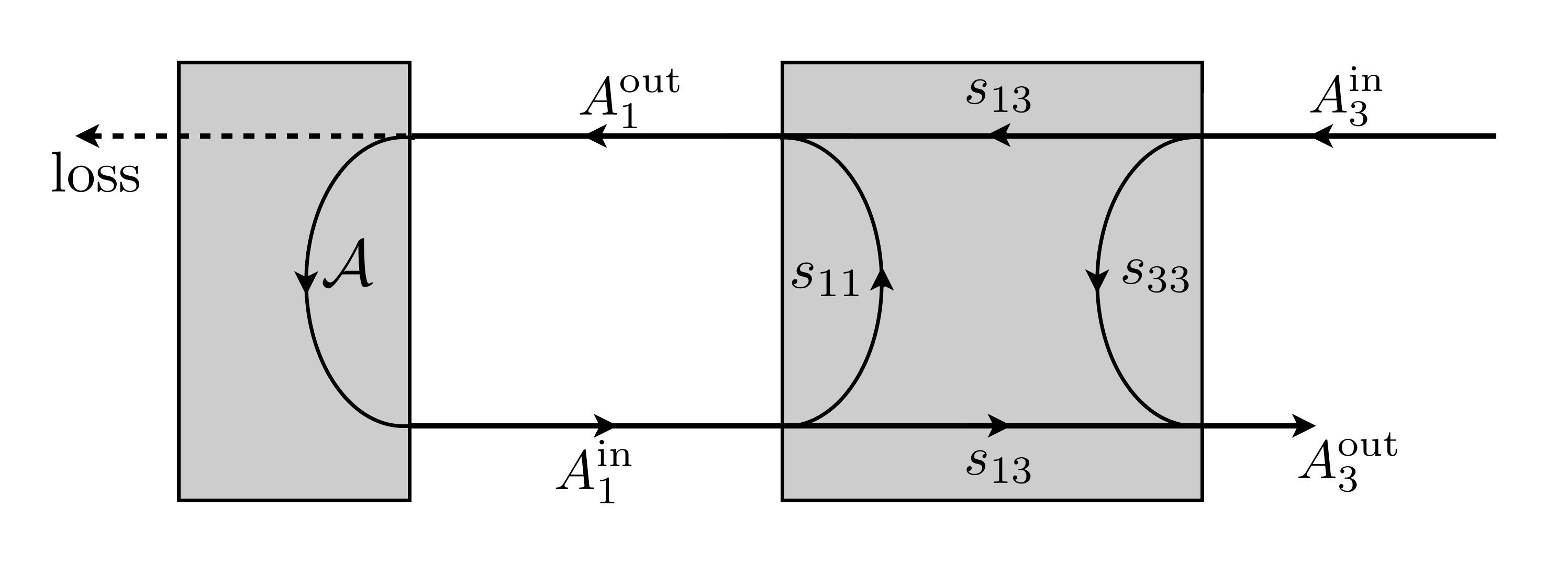}}
\caption{
 A three-body wave arriving from large hyperradius $R$ with amplitude
 $A^{\rm in}_3$ in the triatomic channel can follow various
 pathways before it either returns to large $R$, or gets lost at $R=0$
 by turning into an atom and a deep dimer.
There is a formal analogy with a Fabry-Perot interferometer with two
partially reflecting mirrors. 
} 
\label{fig:fabry}
\end{figure}

It has been noted~\cite{RevueBraaten,Braaten_rec_T} that the loss peaks can be explained by multiple reflections of the hyperradial wave off the intermediate region leading to a resonant denominator under the integral in Eq.~(\ref{eq:K_T}). This behavior becomes transparent if we observe that the considered three-body inelastic scattering problem is formally analogous to a simple interferometer with two partially reflecting mirrors,
see Fig.~\ref{fig:fabry}.
The first mirror (located at intermediate $R$) 
has reflection and transmission amplitudes
 given by the $2 \times 2$ matrix
$$
\begin{pmatrix}
s_{11} & s_{13}
\\
s_{31} & s_{33}
\end{pmatrix}.
$$
The second mirror (located at $R=0$ within the zero-range model)
has the reflection amplitude $\Ar$ which depends on the three-body parameter and the inelasticity parameter [cf. Eq.~(\ref{eq:Ar=})];
transmission through this mirror corresponds to the three-body loss process,
and happens with probability $1-e^{-4\eta}$.

It then becomes clear that the loss probability is modulated by the interference between the different pathways corresponding to multiple reflections by the two mirrors.
More precisely,
rewriting the
 term $1/(1-s_{11}\Ar)$ in Eq.~(\ref{eq:A1out_vs_A3in}) as
$\sum_{n\geq0} (s_{11}\Ar)^n$,
the $n$-th order term corresponds to the pathway with $n$ reflections by each mirror.
This is the origin of the term
$|1 + (k R_0)^{-2 i s_0} e^{-2\eta} \, s_{11} |^{-2} = |1/(1-s_{11}\Ar)|^2$
in the final expression Eq.~(\ref{eq:K_T}).
This also clarifies
why the dependence of the loss rate on $R_0$ and $\eta$ is known analytically,
$s_{11}$ being independent of these parameters.

The interferometer analogy 
also physically explains why
the energy-dependent event rate constant $K(k)$ has the upper bound
\be
K(k) \leq K_{\rm max}(k) = \frac{64\,\pi^2\,\cos^3\!\phi}{mk^4}.
\label{eq:K_k_max}
\ee
This bound is a manifestation of the fact that the loss happens through a single Efimovian channel at short distance (channel $1$),
which is coupled to a single large-distance channel (channel $3$).
The bound is 
reached when a perfect destructive interference leads to $A_3^{\rm out}=0$. The lost flux $\varphi_{\rm loss}$ is then equal to the incoming flux from infinity $|A_3^{\rm in}|^2$ (which is determined by a projection of the incoming plane wave, as we have seen).
After thermal averaging, Eq.~(\ref{eq:K_k_max}) implies
\be
K< K_{\rm max} = \frac{32\,\pi^2\,\cos^3\!\phi\ \hbar^5}{m^{3} (k_B T)^{2}},
\label{eq:K_T_max}
\ee
where the inequality is now strict  since $K(k)=K_{\rm max}(k)$ cannot hold for all $k$
(and we restored the $\hbar$ dependence).

As a side remark we note that in the case where the B particles are
fermionic,
with $\mB/\mA \in (13.60696\ldots;75.99449\ldots)$ so that the Efimov effect occurs in the total angular momentum $L=1$ subspace but not in higher $L$ subspaces~\cite{Efimov73,Petrov3fermions,castin_tignone},
the bounds Eqs.~(\ref{eq:K_k_max}) and (\ref{eq:K_T_max}) are multiplied by 3,
due to the 3 Efimovian sectors with angular momentum projections $M=-1$, 0 and $+1$. 

Let us now briefly address the low-energy limit.
When $k a \to 0^-$, $|s_{11}|\to 1$, i.e., the first mirror becomes nearly perfectly reflecting.
If $\eta$ is small enough, the second mirror is also of good quality, and the finesse of the interferometer becomes sufficiently high to observe resonances in the  variation of the loss rate with scattering length.
The resonances occur when $a$ is such that there exists an Efimov trimer of vanishing energy.
This happens when $s_{11}\Ar \approx 1$, i.e. when the interfering pathways nearly lead to a divergence of
 $(1-s_{11}\Ar)^{-1}$~\cite{Efimov1979}.
This naturally leads to peaks in the loss rate constant $K$ {\it versus} magnetic field, which remain visible and approximately unshifted after thermal averaging
at low enough temperature.

Finally, let us discuss 
the temperature-dependence of the loss rate
at the unitary limit in case {\it i} (see also \cite{WangNJP2011}).
From Eqs.~(\ref{eq:K_T}) and (\ref{eq:s11_unitary}),
one finds
a qualitatively different behavior depending on the order of magnitude of
$|s_{11}^\infty|=e^{-\pi s_0}$.
When
the mass ratio $m_B/m_A$ is sufficiently large, as in the Li-Rb mixture discussed above, $e^{-\pi s_0}$ is small compared to unity, so that $s_{11}^\infty$ can be neglected to a good approximation in Eq.~(\ref{eq:K_T}), leading to $K \approx (1 - e^{-4\eta}) K_{\rm max} \propto 1/T^2$.
In the language of the interferometer, this means that the first mirror is almost transparent, so that interference effects are negligible.
The situation changes
when   $m_B/m_A$ is not large,
{\it e.~g.} for the mixtures  $^6$Li-$^7$Li,
$^{85}$Rb-$^{87}$Rb,
or $^{40}$K-$^{87}$Rb
(for which non-overlapping Feshbach resonances
are indeed available~\cite{FeshbachRMP2010}).
Then, there is significant reflection from the first mirror, and interference effects lead to noticeable log-periodic oscillations of $K\,T^2$ as a function of $T$.
To observe an entire period of these oscillations experimentally is challenging,
since this would require changing $T$ by the large factor $e^{2\pi/s_0}$. However, some $T$-dependence of $K\,T^2$ may be detectable.

\subsection{Relation with previous work}\label{sec:comment}

Let us comment on the similarities and differences between the approach presented here (and introduced in Ref.~\cite{UnitaryBoseGas} in the BBB case)
and previous work on three-boson scattering within the zero-range theory~\cite{Efimov1979,RevueBraaten,Braaten_rec_T}.
Formally, 
it is possible to derive
Eq.~(\ref{eq:K_T}) starting from the $S$-matrix formalism of Refs.~\cite{Efimov1979,RevueBraaten,Braaten_rec_T}. 
However, the present approach is physically more transparent.
A key point is that we directly constructed the relevant large-distance triatomic channel and its hyperangular wavefunction $\Phi_3(\Oo)$, cf.
Eqs.~(\ref{eq:Psi1_smallR},\ref{eq:Psi1_largeR}).
To this end, we considered the wavefunction $\Psi_1$, corresponding to a triatomic flow injected at the origin. 
The idea of this wavefunction $\Psi_1$  was already present in Ref.~\cite{Efimov1979}
where Efimov introduced the concept of $s$ matrix connecting short-distance with large-distance hyperradial motion.
However Ref.~\cite{Efimov1979} focused mainly on the case of negative total energy, where the only open channel at large distances corresponds to motion
of an atom relative to a weakly bound dimer,
while triatomic motion is energetically forbidden.
Braaten and Hammer \cite{RevueBraaten} added a triatomic large-distance channel
charaterized by a wave function independent of the hyperangles.
The resulting  $s$ matrix provides a suitable framework for studying the three-body recombination in a Bose gas with finite $a$ in the zero-temperature limit. 
Then, 
in Ref.~\cite{Braaten_rec_T} this formalism was generalized to finite temperatures by using a complete set of long-distance channels
with hyperangular wavefunctions
defined by hyperspherical harmonics.
A conceptual difficulty with this construction is that these large-distance channels only decouple for $R\gg|a|$, so that one has to formally work with a finite $a$.
In contrast, one would expect physically a smooth dependence on $1/a$ in the interval $(-\infty;0]$.
This expectation is confirmed by the present construction.
The shape of the function $\Phi_3(\Oo)$ depends only on $ka$ and interpolates between two limits: for small $k|a|$ it is a constant independent of $\Oo$ and for large $k|a|$ it tends to $\phi_{is_0}(\Oo)$, the asymptotic region in this case being defined by $R\gg 1/k$. Incidentally, this illustrates the breakdown of the adiabatic hyperspherical approximation for $k|a| \sim 1$.



\section{Calculation of the function $s_{11}$} \label{sec:s11}

Our computational method for the universal function $s_{11}(k a_{\rm AB},ka_{\rm BB})$ consists of finding the three-body wavefunction $\Psi_1$ defined by Eqs.~(\ref{eq:Psi1_smallR}) and (\ref{eq:Psi1_largeR}) numerically. We consider case {\it ii}, while case {\it i} is treated similarly modulo the simplifications
listed at the end of this section. Analytic results for $s_{11}$ near the unitary limit are presented in Appendix~\ref{app:analy}.

The Schr\"odinger equation together with the two-body contact condition for the wave function  $\Psi_1(\RR)$ can be reduced to a set of coupled integral Skorniakov and Ter-Martirosian (STM) equations for the functions $f_{\rm AB}$ and $f_{\rm BB}$ defined by
\bea
f_{\rm AB}({\bf x}) & = & 4\pi\,\lim_{y\to0} y\, \Psi_1,
\nonumber
\\
f_{\rm BB}({\bf X}) & = & 4\pi\,\lim_{Y\to0} Y\, \Psi_1,
\label{eq:def_fAB}
\eea
where $(\xx,\yy)$ are the Jacobi coordinates given in Eqs.~(\ref{eq:def_xy}),
while $(\XX,\YY)$ is another set of Jacobi coordinates defined by
\begin{equation}\label{eq:def_XY}
\renewcommand\arraystretch{1.5}
\begin{array}{r@{}l}
    \cos\theta \ \XX &{}= \rr_2 - (\rr_1 + \rr_3)/2,\\
    2 \sin\theta\ \YY &{}= \rr_3 - \rr_1.    
\end{array}
\end{equation}
The STM equations conserve the angular momentum and, in the considered regime where the mass ratio is not very large, the Efimovian solution appears only in the channel with zero angular momentum, so that $f_{ij}({\bf x})=f_{ij}(x)$.  For this reason the contribution of higher angular momentum channels to the loss 
vanishes in the zero-range limit,
in contrast to the three-body recombination into a weakly-bound state for positive $a$ when $ka$ is not small \cite{Braaten_rec_T}. For more details on the derivation of the STM equations in the general case of different masses and scattering lengths see, for example, Ref.~\cite{PetrovLesHouches2010}.
The asymptotic behavior of $\Psi_1$ at small hyperradius given in
Eq.~(\ref{eq:Psi1_smallR}) translates into
\be
\begin{bmatrix}
f_{\rm AB}(x)
\\
f_{\rm BB}(x)
\end{bmatrix}
\underset{x\to0}{\simeq}
\begin{pmatrix}
C_{\rm AB}
\\
C_{\rm BB}
\end{pmatrix}
\frac{(kx)^{is_0}+s_{11}(kx)^{-is_0}}{x},
\label{eq:f_asym_smallx}
\ee
where $C_{\rm AB}$ and $C_{\rm BB}$ are numerical coefficients given in App.~\ref{app:Cij}.
At large $x$ both functions $f_{\rm AB}(x)$ and $f_{\rm BB}(x)$ should represent outgoing waves:
\be
\begin{bmatrix}
f_{\rm AB}(x)
\\
f_{\rm BB}(x)
\end{bmatrix}
\underset{x\to\infty}{\propto}
\frac{\exp(ikx)}{x^{3/2}}.
\label{eq:f_asym_largex}
\ee

We mention that
at the unitary limit, the solution is known analytically and $f_{ij}$ are expressed in terms of the outgoing Hankel function
\be
\begin{bmatrix}
f_{\rm AB}(x)
\\
f_{\rm BB}(x)
\end{bmatrix}
=
\begin{pmatrix}
C_{\rm AB}
\\
C_{\rm BB}
\end{pmatrix}
\frac{\sinh(s_0\pi)}{e^{s_0\pi}}\,2^{i s_0}\Gamma(1+is_0)
\frac{H^{(1)}_{i s_0}(kx)}{x},
\label{eq:f_Bessel}
\ee
which, matched with Eq.~(\ref{eq:f_asym_smallx}), gives the limiting expression Eq.~(\ref{eq:s11_unitary}).

In order to calculate $s_{11}$ for arbitrary $(k a_{\rm AB},ka_{\rm BB})$ we switch to momentum representation where the STM equations write\begin{widetext}
\begin{equation}\label{eq:STM}
\renewcommand\arraystretch{1.8}
\begin{array}{l@{}l}
&{}(\sqrt{p^2-k^2  - i 0^+}-1/a_{\rm AB})f_{\rm AB}-\hat{L}_{k^2,\phi}f_{\rm AB}-\hat{L}_{k^2,\theta}f_{\rm BB}= 0\\
&{}(\sqrt{p^2-k^2 - i 0^+}-2\sin\theta/a_{\rm BB})f_{\rm BB}-2\hat{L}_{k^2,\theta}f_{\rm AB}= 0,
\end{array}
\end{equation}
where $f_{ij}(p)=\int f_{ij}( x)\exp(-i{\bf px}) d^3x$,
\begin{equation}\label{eq:L}
\hat{L}_{k^2,\alpha}f(p)=\frac{1}{\pi\sin2\alpha}\int_0^\infty 
dp'\,
f(p')
\frac{p'}{p} \ln\left(\frac{p'^2+2pp'\sin\alpha+p^2-k^2\cos^2\alpha - i
    0^+ }{p'^2-2pp'\sin\alpha+p^2-k^2\cos^2\alpha - i
    0^+ }\right),
\end{equation}
$\phi$ and $\theta$ are the mass angles defined in
Eqs.~(\ref{eq:phi}) and (\ref{eq:def_theta}), and the inclusion of a small positive $0^+$ indicates that $k$ is slightly shifted into the upper complex halfplane thus fixing the branch cuts of the logarithm and square root and ensuring the presence of only the outgoing wave (\ref{eq:f_asym_largex}) in the solution. The boundary condition (\ref{eq:f_asym_smallx}) now reads
\be
\begin{bmatrix}
f_{\rm AB}(p)
\\
f_{\rm BB}(p)
\end{bmatrix}
\underset{p\to\infty}{\simeq}
\begin{pmatrix}
C_{\rm AB}
\\
C_{\rm BB}
\end{pmatrix}\frac{2\pi^2 s_0}{\sinh(\pi s_0/2)}
\left[\frac{1}{\Gamma(1-is_0)}\frac{(p/k)^{-is_0}}{p^2}+s_{11}\frac{1}{\Gamma(1+is_0)}\frac{(p/k)^{is_0}}{p^2}\right].
\label{eq:f_asym_largep}
\ee

We then perform a complex scaling transformation.
We use the fact that $s_{11}$ depends only on the products
$a_{\rm AB}k$ and $a_{\rm BB}k$ and replace in Eq.~(\ref{eq:STM})
$k\rightarrow i$, $a_{ij}\rightarrow -ia_{ij}k$. 
Equivalently, 
this can be done by rotating the integration contour over $p'$ in Eq.~(\ref{eq:L}) to the negative imaginary axis.
The obtained STM equations,
\begin{equation}\label{eq:STMrescaled}
\renewcommand\arraystretch{1.8}
\begin{array}{l@{}l}
&{}(\sqrt{p^2+1}-i/a_{\rm AB}k)f_{\rm AB}-\hat{L}_{-1,\phi}f_{\rm AB}-\hat{L}_{-1,\theta}f_{\rm BB}=0,\\
&{}(\sqrt{p^2+1}-2i\sin\theta/a_{\rm BB}k)f_{\rm BB}-2\hat{L}_{-1,\theta}f_{\rm AB}=0,
\end{array}
\end{equation} 
do not contain singularities in the kernels and can be efficiently solved numerically. We deduce $s_{11}$ from matching the solution with the asymptotic form
\be
\begin{bmatrix}
f_{\rm AB}(p)
\\
f_{\rm BB}(p)
\end{bmatrix}
\underset{p\to\infty}{\propto}
\begin{pmatrix}
C_{\rm AB}
\\
C_{\rm BB}
\end{pmatrix}
\left[p^{-2-is_0}+s_{11}e^{\pi s_0}\frac{\Gamma(1-is_0)}{\Gamma(1+is_0)}p^{-2+is_0}\right]
\label{eq:f_asym_largeprescaled}
\ee
\end{widetext}
for any given set $a_{\rm AB}k$ and $a_{\rm BB}k$. More precisely, we use the following procedure. We introduce new unknown functions $g_{\rm AB}(p)$ and $g_{\rm BB}(p)$ such that $f_{ij}(p) \propto [p^{-2-is_0}+g_{ij}(p)]$ [the proportionality coefficients follow from Eq.~(\ref{eq:f_asym_largep}) but are irrelevant for the determination of $s_{11}$] and require that for large momenta $g_{ij}(p)\propto p^{-2+is_0}$. Equations~(\ref{eq:STMrescaled}) become a set of linear inhomogeneous equations for $g_{AB}$ and $g_{BB}$, which we solve by discretizing $p$ and inverting the corresponding discrete analog of the operator in the left hand side of Eqs.~(\ref{eq:STMrescaled}). The final result is given by $s_{11}=e^{-\pi s_0}\Gamma(1+is_0)/\Gamma(1-is_0)\lim_{p\rightarrow\infty} g_{ij}(p)p^{2-is_0}$ independent of the choice $g_{\rm AB}$ or $g_{\rm BB}$.

Case {\it i} is obtained by setting $f_{\rm BB}=C_{\rm BB}=a_{\rm BB}=0$ in the above analysis and omitting the second lines in the STM equations (\ref{eq:STM}) and (\ref{eq:STMrescaled}). Various properties of the ABB system in case {\it i} have been studied by using the STM equation derived in this case from the effective field theory~\cite{HelfrichHammerPetrov}. The STM equations in both cases {\it i} and {\it ii} have been employed 
in a
recent study of the single-particle momentum distribution of heteronuclear Efimov trimers~\cite{Yamashita2013spm}.

Another point which is useful to mention is that Braaten {\it et al.}~\cite{Braaten_rec_T} have developed a slightly different method for calculating $s_{11}$ in the case of three identical bosons. They also use the STM equation but deduce $s_{11}$ from solving the atom-dimer scattering problem for positive $a$ above the dimer break-up threshold. We notice that their approach works well for small $ka>0$ whereas ours is optimal for large $ka$, either positive or negative. In fact, $s_{11}(-a_{\rm AB}k,-a_{\rm BB}k)=e^{-2\pi s_0}/s_{11}^*(a_{\rm AB}k,a_{\rm BB}k)$. This relation follows from Eq.~(\ref{eq:f_asym_largeprescaled}) and the observation that if $f_{ij}$ is the solution of Eqs.~(\ref{eq:STMrescaled}) for $a_{ij}$, then $f_{ij}^*$ is the solution for $a_{ij}=-a_{ij}$.

\section{Summary and outlook}

In this paper we have obtained the finite-temperature three-body loss rate constant within the zero-range theory for the ABB system, with $a_{\rm AB}<0$, and no BB interaction (case {\it i}) or $a_{\rm BB}<0$ (case {\it ii}). For a given mass ratio, we expressed the rate constant in terms of temperature $T$, three-body parameter $R_0$, inelasticity parameter $\eta$, and a universal function $s_{11}$ that depends on $ka_{\rm AB}$,
as well as on $ka_{\rm BB}$ in case {\it ii},
where $k$ is the three-body collision momentum.
We developed a numerical method based on complex scaling for computing the function $s_{11}$ and perform this calculation for two experimentally relevant cases:
$^6$Li-$^{133}$Cs-$^{133}$Cs
and
$^6$Li-$^{87}$Rb-$^{87}$Rb.
The knowledge of $s_{11}$ reduces the problem of computing the loss rate to a simple thermal averaging integral over $k$ for any desired $T$, $R_0$ and $\eta$.
We expect that these results, combined with experimental data, will be useful for precise tests of universality and determinations of $R_0$ and $\eta$.
For $^6$Li-$^{133}$Cs-$^{133}$Cs we find that inclusion of the CsCs interaction leads to a significant enhancement of the loss rate constant. 
This is in spite of the fact that the scaling factors in cases {\it i} and {\it ii} are very close for this large mass ratio. The enhanced loss rate is likely to be explained by an enhanced probability to find two Cs atoms close to each other in the CsCs-interacting case.

In deriving the loss rate coefficient we explicitly constructed the hyperangular wave function $\Phi_3(\Oo)$ corresponding to the long-distance three-atom channel which is connected by a unitary two-by-two matrix with the Efimovian wave at small hyperradii.
The corresponding $S$-matrix formalism smoothly connects
with the exactly solvable unitarity limit.
The three-body loss problem reduces to a simple Fabry-Perot interferometer with two partially reflecting mirrors.

Three-body systems at zero total energy are known to be analytically solvable in the zero-range approximation if there is only one relevant scattering length (three identical bosons \cite{Macek1,Macek2,Gogolin2008aso,MoraCRAS,PetrovLesHouches2010}, ABB system in case {\it i} \cite{HelfrichHammerPetrov}, ABB system with fermionic B particles \cite{Petrov3fermions}). A theoretical challenge that we can formulate now is to generalize these approaches to the ABB system in case {\it ii} at vanishing total energy. In particular, it would be interesting to obtain an analytic expression for the loss rate constant at zero temperature (the dash-dotted line in Fig.~\ref{fig:KLiCsCs}). 

\acknowledgments
We thank A.~Bulgac, F.~Chevy, A.~Grier, R.~Grimm, and E. Kuhnle for fruitful discussions and acknowledge support from the IFRAF Institute and the Institute for Nuclear Theory during the program {\it Universality in Few-Body Systems: Theoretical Challenges and New Directions, INT-14-1}. The research leading to these results has received funding from the European Research Council under European Community's Seventh Framework Programme (FR7/2007-2013 Grant Agreement no.341197),
ERC (grant `Thermodynamix'), and IFRAF-NanoK (grant `Atomix').

{\it Note:}
During completion of this manuscript, we became aware of a related work~\cite{ZinnerHetero}
that uses the adiabatic hyperspherical approximation and also finds a strong effect of the BB interaction.

\appendix

\section{Basic properties of the $s$ matrix} \label{app:s_matrix}

Here we derive some basic properties of the $s$ matrix that were stated and used in the main text.

Let us first give a proper definition of the $s$ matrix.
In addition to the state $\Psi_1$ defined above,
let us define the state $\Psi_m$
for $m\geq3$
as the solution of the Schr\"odinger equation with the two-body contact condition Eqs.~(\ref{eq:schro},\ref{eq:2body_cc}) 
whose asymptotic behavior contains an incoming wave
 of unit amplitude
in channel $m$
and a purely outgoing wave
in the other channels;
the coefficients of the outgoing waves then define the column $(s_{nm})_{n\in\Cr}$ of the $s$ matrix:
\begin{numcases}{}
      \Psi_m   \underset{R\to0}{\simeq}  s_{1 m}\,\psi_1^{\rm out} & \label{eq:psi_n_0} 
\nonumber
\\
      \Psi_m   \underset{R\to\infty}{\simeq}  \psi_m^{\rm in} + \sum_{n\geq3} s_{n m}\,\psi_n^{\rm out}.& \label{eq:psi_n_inf}
 \end{numcases}

A useful lemma is that for an arbitrary state $\varPsi$
solving Schr\"odinger's equation with the two-body contact condition Eqs.~(\ref{eq:schro},\ref{eq:2body_cc}),
the in- and out-amplitudes [defined in Eqs.~(\ref{eq:psi_ampl_0},\ref{eq:psi_ampl})]
are constrained by
\be
\sum_{n\in\Cr} |A_n^{\rm in}|^2 = \sum_{n\in \Cr} |A_n^{\rm out}|^2.
\label{eq:fluxA}
\ee
This follows from the conservation of probability.
More precisely, the
flux through a hypersurface $\Sr$ [defined as in Eqs.~(\ref{eq:def_J},\ref{eq:def_phi_loss})] is independent of $\Sr$.
Taking for $\Sr$ a hypersphere of very small or very large radius,
the flux respectively equals
$|A_1^{\rm out}|^2-|A_1^{\rm in}|^2$
or
$\sum_{n\geq3} ( |A_n^{\rm out}|^2-|A_n^{\rm in}|^2 )$.

A useful consequence is that if 
$\varPsi$ and
$\varPsi'$ are two solutions of
Eqs.~(\ref{eq:schro},\ref{eq:2body_cc})
with the same in-amplitudes, then  their out-amplitudes are also equal.
This follows from applying the above lemma to $\varPsi-\varPsi'$.

The linear relation between out- and in-amplitudes Eq.~(\ref{eq:linS})
then follows by noting that
$\varPsi$ and $\sum_{n\in\Cr} A_n^{\rm in}\,\Psi_n$ have equal in-amplitudes
and hence equal out-amplitudes.


Let us now check that the matrix $s_{nm}$ is unitary.
Physically this comes again from the conservation of probability.
More precisely,  applying the lemma Eq.~(\ref{eq:fluxA}) to the state $\varPsi=\Psi_n+\alpha\,\Psi_{n'}$ where $n\neq n'$ and $\alpha$ is an arbitrary complex number, we get
\bea
1+|\alpha|^2 &=&
\sum_{m\in\Cr} | s_{mn}+\alpha \, s_{m n'} |^2.
\label{eq:alpha}
\eea
Taking $\alpha=0$ yields
\be
\sum_m |s_{mn}|^2=1.
\label{eq:s_U_1}
\ee
Thus (\ref{eq:alpha}) simplifies to 
${\rm Re} ( \alpha \sum_m
s_{mn}^* s_{m n'})=0$ for any $\alpha$, which implies
\be
\sum_m
s_{mn}^* s_{m n'}=0.
\label{eq:s_U_2}
\ee

Let us now check that $s_{nm}$ is symmetric. 
Physically, this follows from time-reversal invariance. More precisely, we consider the state $\Psi_n^*$, 
and notice that it has the same in-amplitudes than
$\sum_{m\in\Cr} s_{m n}^*\,\Psi_m$,
hence also the same out-ampltiudes.
This yields
 $s\cdot s^* = \mathbf{1}$,
and hence, since $s$ is unitary,
 $s =s^T$.

\section{Determination of $s_0$, $C_{\rm AB}$ and $C_{\rm BB}$} \label{app:Cij}

The STM equations can be used to calculate $s_0$ and the ratio of $C_{\rm AB}$ and $C_{\rm BB}$. 
To this end, it suffices to consider the scale-invariant case where 
the interaction is at the unitary limit
and the energy is zero.

Let us first consider case {\it ii}.
A solution of the STM equations~(\ref{eq:STM}),
for
 $k=0$, 
is given by the ansatz \cite{Yamashita2013spm}
\be
\begin{bmatrix}
f_{\rm AB}(x)
\\
f_{\rm BB}(x)
\end{bmatrix}
=
\begin{pmatrix}
C_{\rm AB}
\\
C_{\rm BB}
\end{pmatrix}
\, \frac{p^{i s_0}}{p^2}.
\nonumber
\ee
Indeed, we have $\hat{L}_{0,\alpha}p^{is_0}/p^2=\lambda(s_0,\alpha)p^{is_0}/p$, where 
the function $\lambda$ is defined by Eq.~(\ref{eq:def_lambda}).
The STM equations then become a 2$\times$2 homogeneous system of linear algebraic equations for $C_{\rm AB}$ and $C_{\rm BB}$:
\begin{equation}\label{eq:mat_CAB}
\begin{pmatrix}
1-\lambda(s_0,\phi)&-\lambda(s_0,\theta)\\
-2\lambda(s_0,\theta)&1
\end{pmatrix}
\begin{pmatrix}
C_{\rm AB} \\ C_{\rm BB}
\end{pmatrix}
=0.
\end{equation}
The requirement that the determinant of the 2$\times$2 matrix in Eq.~(\ref{eq:mat_CAB}) vanish gives the implicit equation for $s_0$ Eq.~(\ref{eq:s_with_BB}).
The ratio between $C_{\rm AB}$ and $C_{\rm BB}$ is then fixed by 
$C_{\rm BB}=2\lambda(s_0,\theta)C_{\rm AB}$
 and we can choose these coefficients to be real.

In case {\it i},
one should formally set $C_{\rm BB}=0$ in the above analysis.

Finally we note that, even though
all that is used
for the computation of $s_{11}$ in Sec.~\ref{sec:s11}
is the ratio of $C_{\rm AB}$ and $C_{\rm BB}$,
their absolute values can also be determined, thanks to the expression of the normalized wavefunction $\phi_{is_0}(\Oo)$
given in Appendix~\ref{app:wavefunc}.
Indeed, Eqs.~(\ref{eq:Psi1_smallR},\ref{eq:def_fAB},\ref{eq:phi_norm},\ref{eq:phi_alpha}) yield straightforwardly
\be
\begin{pmatrix}
C_{\rm AB}
\\
C_{\rm BB}
\end{pmatrix}
=
\,4\pi\sinh\left(s_0\frac{\pi}{2}\right)
\,\sqrt{\frac{m}{2s_0}}\,
\begin{pmatrix}
\Cr_{\rm AB}
\\
\Cr_{\rm BB}
\end{pmatrix},
\ee
the absolute values of $\Cr_{\rm AB}$ and $\Cr_{\rm BB}$ being determined by Eq.~(\ref{eq:CAB_norm}).

\section{Normalised wavefunction and contact parameters of a heteronuclear Efimov trimer}
\label{app:wavefunc}

In this Appendix, we consider the three-body bound states,
{\it i.e.}
the negative-energy solutions of the zero-range model [defined by the three-body Schr\"odinger equation Eq.~(\ref{eq:schro}), the two-body contact condition Eq.~(\ref{eq:2body_cc}), and the three-body contact condition Eq.~(\ref{eq:3body_cc})],
 at the unitary limit Eq.~(\ref{eq:unitary_limit})
and without losses ($\eta=0$).
 We provide the analytical expressions of their normalised wavefunction and their contact parameters.
We
 consider case {\it ii}
 throughout this Appendix.
The content of this Appendix
also applies to case {\it i} provided
$\Cr_{\rm BB}$ is formally set to zero.

\subsection{Wavefunction}

The trimer's wavefunction writes~\cite{Efimov73}
\be
\psi( \RR) = \frac{F(R)}{R^2}\,\phi_{i s_0}(\Oo)
\ee
where the hyperradial part is proportionnal to a Bessel function
\be
F(R) = \Nr\,K_{i s_0}(\sqrt{m|E|}\, R)
\ee
and the hyperangular part writes
\be
\label{eq:phi_norm}
\phi_{i s_0}(\Oo) = (1+ P)\,\Cr_{\rm AB} \frac{\varphi(\alpha)}{\sin\alpha\cos\alpha}
+
\Cr_{\rm BB} \frac{\varphi(\beta)}{\sin\beta\cos\beta}
\ee
where
\be
\varphi(\alpha) = \sinh\left[ s_0\,\left(\frac{\pi}{2}-\alpha\right)\right],
\label{eq:phi_alpha}
\ee
$\alpha$ and $\beta$ being hyperangles
 defined by
\bea
x&=& R\,\cos\alpha
\nonumber
\\
y&=& R\,\sin \alpha
\nonumber
\eea
and
\bea
X&=& R\,\cos\beta
\nonumber
\\
Y&=& R\,\sin \beta,
\nonumber
\eea
the Jacobi coordinates $(x,y)$ and $(X,Y)$ being defined in Eqs.~(\ref{eq:def_xy}) and (\ref{eq:def_XY}).

Imposing the two-body contact conditions Eqs.~(\ref{eq:2body_cc}) yields that $\Cr_{\rm AB}$ and $\Cr_{\rm BB}$ satisfy the same matrix equation Eq.~(\ref{eq:mat_CAB}) than 
$C_{\rm AB}$ and $C_{\rm BB}$,
which again yields Eq.~(\ref{eq:s_with_BB}) for $s_0$,
as well as
\be
\Cr_{\rm BB}=2\lambda(s_0,\theta)\Cr_{\rm AB}.
\label{eq:CrAB_ratio}
\ee

The spectrum is
\be
E_n = - \frac{1}{m R_0^2}\,4\,e^{2\,{\rm arg}\,\Gamma(1+i s_0) / s_0}\, e^{-n \, 2\pi/s_0}, \ \ n\in\mathbb{Z},
\label{eq:Eefi}
\ee
as obtained by imposing the three-body contact condition Eq.~(\ref{eq:3body_cc}). 


The normalization of the wavefunction can be done analytically, 
generalizing~\cite{CastinWerner_nk_trimer} to the heteronuclear case.
For the hyperangular part, we take $(\phi_{i s_0}|\phi_{i s_0}) = 1$ for the hyperangular scalar product introduced in Eq.~(\ref{eq:def_ps}).
This leads to
\begin{equation}
(2\Cr_{\rm AB}^{\phantom{ab}2}+\Cr_{\rm BB}^{\phantom{ab}2})Q(\pi/2)+2\Cr_{\rm AB}^{\phantom{ab}2}Q(\phi)+4\Cr_{\rm AB}\Cr_{\rm BB}Q(\theta)=1,
\label{eq:CAB_norm}
\end{equation}
where
\begin{widetext}
\be
\label{eq:F}
Q(\alpha)= \frac{8\pi^2}{s_0 \sin2\alpha} \left[ \pi\cosh\left(s_0 \frac{\pi }{2}\right)\sinh(s_0\,\alpha )
-2\alpha \sinh\left(s_0\frac{\pi}{2}\right)\cosh(s_0\,\alpha ) \right]
\ee
\end{widetext}
 and 
$$Q(\pi/2) \equiv \lim_{\alpha\to\pi/2}Q(\alpha) = 4\pi^2[\sinh(s_0\pi)/s_0-\pi].$$

In order to normalize the hyperradial wavefunction, 
it is convenient to consider that the three particles are subject to an external harmonic trapping potential of vanishing frequency~\cite{CastinWerner_nk_trimer}, so that one can impose $\int d^3r_1 d^3r_2 d^3r_3 |\Psi(\rr_1,\rr_2,\rr_3)|^2=1$, where $\Psi$ equals $\psi(\RR)$ times a center-of-mass wavefunction
$\psi_{\rm CM}(\CC)$
normalized to $\int d^3C |\psi_{\rm CM}(\CC)|^2 = 1$. 
Due to the Jacobian Eq.~(\ref{eq:jacobian}),
this gives $\int_0^\infty dR\,R\,|F(R)|^2 = \cos^{-3}\phi$. 
The integral over $R$ has
a known expression~\cite{Gradstein}, which finally yields
\be
\Nr = \sqrt{\frac{2 m |E| \,\sinh(s_0\pi)}{s_0\pi\,\cos^3\!\phi}}.
\ee

\subsection{Contact parameters}

From this normalized wavefunction, it is
straightforward to deduce the contact parameters of an Efimov trimer, {\it i.~e.}
the partial derivatives of its energy with respect to the scattering
lengths
taken at fixed three-body parameter.
Indeed,
as shown in Ref.~\cite{WernerCastinRelationsBosons},
 $(\partial E/\partial a_{ij})_{R_0}$ is proportional to the
norm of the function $A_{ij}$ appearing in the 2-body
contact-condition Eq.~(\ref{eq:2body_cc}).
This yields
\begin{widetext}
\bea
\left. \frac{\partial E}{\partial (-1/a_{\rm AB})} \right|_{R_0}
&=&
\Cr_{\rm AB}^{\phantom{ab}2}
\
\sqrt{\frac{|E|}{m}}
\
16\pi^3\,\frac{\tanh(s_0\pi)\,\sinh^2\left(s_0 \pi/2\right)}{ s_0}
\label{eq:dE_daAB}
\\
\left. \frac{\partial E}{\partial (-1/a_{\rm BB})} \right|_{R_0} &=& 
\Cr_{\rm BB}^{\phantom{ab}2}
\
\sqrt{\frac{|E|}{m}}
\
16\pi^3
\,\frac{\tanh(s_0\pi)\,\sinh^2\left(s_0 \pi/2\right)}{ s_0}
\
\sin\theta.
\label{eq:dE_daBB}
\eea
This generalizes to the heteronuclear case the result obtained in~\cite{CastinWerner_nk_trimer} for three identical bosons.

\section{Expansion of $s_{11}$ around the unitary limit}
\label{app:analy}

This Appendix concerns leading-order corrections to $s_{11}$ near the unitary limit.
We start by stating the results.
In case {\it ii}, when $k|a_{\rm AB}|$ and $k|a_{\rm BB}|$ tend to $\infty$,
we have the expansion
\begin{equation}\label{eq:s11_unitary_plus}
s_{11}\approx s_{11}^\infty \, \left[1
 - 
16\,\pi^3\,\tanh(s_0\pi)\,\sinh^2\!\left(s_0\frac{\pi}{2}\right)
\left(\frac{\Cr_{\rm AB}^{\phantom{ab}2}}{a_{\rm AB}k}+\frac{\Cr_{\rm BB}^{\phantom{ab} 2}\,\sin\theta}{a_{\rm BB}k}\right)\right],
\end{equation}
\end{widetext}
where $s_{11}^\infty$ is given by Eq.~(\ref{eq:s11_unitary}) and the coefficients $\Cr_{\rm AB}$ and $\Cr_{\rm BB}$ are derived from
Eqs.~(\ref{eq:CrAB_ratio},\ref{eq:CAB_norm}).
In case {\it i}, the expansion 
for $k|a_{\rm AB}|\to\infty$
is given by formally setting $\Cr_{\rm BB}=0$ in Eqs.~(\ref{eq:s11_unitary_plus},\ref{eq:CAB_norm}).

We present the derivation in case {\it ii}
(case {\it i} is treated similarly with obvious simplifications).
We employ the same complex scaling procedure
than in Sec.~\ref{sec:s11}, now  at the level of the Schr\"odinger
equation. Namely,
we assume that the function $Z \mapsto \Psi_1(Z\,\RR)$, defined {\it a
  priori} for real positive $Z$, can be analytically continued to the
quadrant $0\leq {\rm Arg}\, Z \leq \pi/2$, and we consider the scaled wavefunction
$\tilde{\Psi}_1(\tilde{\RR}) \equiv \Psi_1(i\,\tilde{\RR} / k)$
where $\tilde{\RR}$ has real coordinates.
The Schr\"odinger equation $-\Delta_{\RR}\Psi_1 = k^2\,\Psi_1$
then becomes
$-\Delta_{\tilde{\RR}}\tilde{\Psi}_1 = -\tilde{\Psi}_1$.
Furthermore, the large-distance behavor $\Psi_1 \propto e^{i k R}$
gives $\tilde{\Psi}_1 \propto e^{-\tilde{R}}$ after analytic continuation.
Finally, because $\Psi_1$ satisfies the two-body contact conditions with scattering lengths $(a_{\rm AB}, a_{\rm BB})$,
$\tilde{\Psi}_1$ satisfies the two-body contact conditions with the scaled scattering lengths 
\be
(\tilde{a}_{\rm AB}, \tilde{a}_{\rm BB}) = (a_{\rm AB}, a_{\rm BB}) \, k / i.
\label{eq:scale_a}
\ee
The short-distance behavor
$\Psi_1 \propto (k R)^{i s_0} + s_{11}\,(k R)^{-i s_0}$
turns into
$\tilde{\Psi}_1 \propto \tilde{R}^{i s_0} + s_{11}\,e^{\pi s_0}\,\tilde{R}^{-i s_0}$,
which we rewrite as
$\tilde{\Psi}_1 \propto (\tilde{R}/\tilde{R}_0)^{-i s_0} - (\tilde{R}/\tilde{R}_0)^{i s_0}$
where $\tilde{R}_0$ can be viewed as an effective (in general complex) three-body parameter, related to $s_{11}$ by $s_{11}\,e^{\pi s_0}=-\tilde{R}_0^{\ph 2 i s_0}$.
Summarizing,
the scaled wavefunction corresponds to a bound trimer state, of fixed energy $-1/m$, with imaginary scattering lengths and complex three-body parameter.

At the unitary limit, the rescaling of the scattering lengths Eq.~(\ref{eq:scale_a}) has no effect, and $\tilde{\Psi}_1$ is the wavefunction of a standard Efimov trimer. 
Hence 
its effective three-body parameter $\tilde{R}_0$ is real, and is related to
its energy, $-1/m$, as in Eq.~(\ref{eq:Eefi}).
This allows to retrieve the expression Eq.~(\ref{eq:s11_unitary})
of $s_{11}^\infty$.

If we move slightly away from the unitary limit by turning on small finite inverse scattering lengths,
 $\tilde{R}_0$ has to shift in such a way that the scaled energy remains fixed. 
But
the partial derivatives of the energy with respect to the inverse scattering lengths are given by the contact parameters, computed in App.~\ref{app:wavefunc}.
The partial derivative of the energy with respect to the three-body parameter, on the other hand, is obtained easily from the simple relation between energy and three-body parameter valid at the scale-invariant point.
This yields the desired result Eq.~(\ref{eq:s11_unitary_plus}).

\section{Three identical bosons} \label{app:homo}

The case of recombination between three bosons in the same internal state,
which was the subject of Ref.~\cite{UnitaryBoseGas},
can be recovered from the present article,
modulo minor modifications given in this Appendix.
One  sets $\mA=\mB=m$ so that $\phi = \theta=\pi / 6$,
and
$a_{\rm AB}=a_{\rm BB}=: a$
(case {\it i} does not exist any more).
In Eq.~(\ref{eq:s}) the right hand side is multiplied by $2$.
For the symmetrization and normalisation of the plane wave, in  Eq.~(\ref{eq:sym_eikR}),
the term $(1+P)/\sqrt{2}$
 is replaced by
$(\sum_{\sigma} P_{\sigma})/\sqrt{3!}$,
where the sum over $\sigma$ now runs over the $3!$ permutations of the three particles.
As a consequence, the right hand side of
Eq.~(\ref{eq:A3in})
(and of its finite-$a$ generalization)
gets multiplied by $\sqrt{3}$.
Furthermore,
in Eq.~(\ref{eq:K_k_vs_loss})
the right hand side contains an additional factor $1/3$.
The additional factors cancel out
in the final
 result for $K$ and Eq.~(\ref{eq:K_T}) is unchanged.

As far as the calculation of the function $s_{11}$ is concerned, Sec.~\ref{sec:s11} remains entirely valid in the case BBB, but gets simplified by observing that $f_{\rm AB}=f_{\rm BB}$ and $C_{\rm AB}=C_{\rm BB}$ which leads to the usual single STM equation. Similarly, all analytic results in Apps.~\ref{app:wavefunc} and \ref{app:analy} reproduce the known BBB ones (in this case $\Cr_{\rm AB}=\Cr_{\rm BB}$, and 
$\partial_{1/a_{\rm AB}} + \partial_{1/a_{\rm BB}}$ should be replaced by $\partial_{1/a}$).
For completeness we also mention the relation between $a_-$ and $R_0$ in this case, $a_- \approx -1.017\, R_0$.

\bibliography{bibliography}

\begin{thebibliography}{55}%
\makeatletter
\providecommand \@ifxundefined [1]{%
 \@ifx{#1\undefined}
}%
\providecommand \@ifnum [1]{%
 \ifnum #1\expandafter \@firstoftwo
 \else \expandafter \@secondoftwo
 \fi
}%
\providecommand \@ifx [1]{%
 \ifx #1\expandafter \@firstoftwo
 \else \expandafter \@secondoftwo
 \fi
}%
\providecommand \natexlab [1]{#1}%
\providecommand \enquote  [1]{``#1''}%
\providecommand \bibnamefont  [1]{#1}%
\providecommand \bibfnamefont [1]{#1}%
\providecommand \citenamefont [1]{#1}%
\providecommand \href@noop [0]{\@secondoftwo}%
\providecommand \href [0]{\begingroup \@sanitize@url \@href}%
\providecommand \@href[1]{\@@startlink{#1}\@@href}%
\providecommand \@@href[1]{\endgroup#1\@@endlink}%
\providecommand \@sanitize@url [0]{\catcode `\\12\catcode `\$12\catcode
  `\&12\catcode `\#12\catcode `\^12\catcode `\_12\catcode `\%12\relax}%
\providecommand \@@startlink[1]{}%
\providecommand \@@endlink[0]{}%
\providecommand \url  [0]{\begingroup\@sanitize@url \@url }%
\providecommand \@url [1]{\endgroup\@href {#1}{\urlprefix }}%
\providecommand \urlprefix  [0]{URL }%
\providecommand \Eprint [0]{\href }%
\providecommand \doibase [0]{http://dx.doi.org/}%
\providecommand \selectlanguage [0]{\@gobble}%
\providecommand \bibinfo  [0]{\@secondoftwo}%
\providecommand \bibfield  [0]{\@secondoftwo}%
\providecommand \translation [1]{[#1]}%
\providecommand \BibitemOpen [0]{}%
\providecommand \bibitemStop [0]{}%
\providecommand \bibitemNoStop [0]{.\EOS\space}%
\providecommand \EOS [0]{\spacefactor3000\relax}%
\providecommand \BibitemShut  [1]{\csname bibitem#1\endcsname}%
\let\auto@bib@innerbib\@empty
\bibitem [{\citenamefont {Braaten}\ and\ \citenamefont
  {Hammer}(2007)}]{RevueBraaten2}%
  \BibitemOpen
  \bibfield  {author} {\bibinfo {author} {\bibfnamefont {E.}~\bibnamefont
  {Braaten}}\ and\ \bibinfo {author} {\bibfnamefont {H.-W.}\ \bibnamefont
  {Hammer}},\ }\href@noop {} {\bibfield  {journal} {\bibinfo  {journal} {Ann.
  Phys.}\ }\textbf {\bibinfo {volume} {322}},\ \bibinfo {pages} {120} (\bibinfo
  {year} {2007})}\BibitemShut {NoStop}%
\bibitem [{\citenamefont {Ferlaino}\ \emph {et~al.}(2011)\citenamefont
  {Ferlaino}, \citenamefont {Zenesini}, \citenamefont {Berninger},
  \citenamefont {Huang}, \citenamefont {N\"agerl},\ and\ \citenamefont
  {Grimm}}]{EfimovReviewGrimm}%
  \BibitemOpen
  \bibfield  {author} {\bibinfo {author} {\bibfnamefont {F.}~\bibnamefont
  {Ferlaino}}, \bibinfo {author} {\bibfnamefont {A.}~\bibnamefont {Zenesini}},
  \bibinfo {author} {\bibfnamefont {M.}~\bibnamefont {Berninger}}, \bibinfo
  {author} {\bibfnamefont {B.}~\bibnamefont {Huang}}, \bibinfo {author}
  {\bibfnamefont {H.-C.}\ \bibnamefont {N\"agerl}}, \ and\ \bibinfo {author}
  {\bibfnamefont {R.}~\bibnamefont {Grimm}},\ }\href@noop {} {\bibfield
  {journal} {\bibinfo  {journal} {Few-Body Syst.}\ }\textbf {\bibinfo {volume}
  {51}},\ \bibinfo {pages} {113} (\bibinfo {year} {2011})}\BibitemShut
  {NoStop}%
\bibitem [{\citenamefont {Gross}\ \emph {et~al.}(2011)\citenamefont {Gross},
  \citenamefont {Shotan}, \citenamefont {Machtey}, \citenamefont {Kokkelmans},\
  and\ \citenamefont {Khaykovich}}]{KhaykovichEfimovCRAS}%
  \BibitemOpen
  \bibfield  {author} {\bibinfo {author} {\bibfnamefont {N.}~\bibnamefont
  {Gross}}, \bibinfo {author} {\bibfnamefont {Z.}~\bibnamefont {Shotan}},
  \bibinfo {author} {\bibfnamefont {O.}~\bibnamefont {Machtey}}, \bibinfo
  {author} {\bibfnamefont {S.}~\bibnamefont {Kokkelmans}}, \ and\ \bibinfo
  {author} {\bibfnamefont {L.}~\bibnamefont {Khaykovich}},\ }\href@noop {}
  {\bibfield  {journal} {\bibinfo  {journal} {C. R. Physique}\ }\textbf
  {\bibinfo {volume} {12}},\ \bibinfo {pages} {4} (\bibinfo {year}
  {2011})}\BibitemShut {NoStop}%
\bibitem [{\citenamefont {Kraemer}\ \emph {et~al.}(2006)\citenamefont
  {Kraemer}, \citenamefont {Mark}, \citenamefont {Waldburger}, \citenamefont
  {Danzl}, \citenamefont {Chin}, \citenamefont {Engeser}, \citenamefont
  {Lange}, \citenamefont {Pilch}, \citenamefont {Jaakkola}, \citenamefont
  {Naegerl},\ and\ \citenamefont {Grimm}}]{GrimmEfimovEvidence}%
  \BibitemOpen
  \bibfield  {author} {\bibinfo {author} {\bibfnamefont {T.}~\bibnamefont
  {Kraemer}}, \bibinfo {author} {\bibfnamefont {M.}~\bibnamefont {Mark}},
  \bibinfo {author} {\bibfnamefont {P.}~\bibnamefont {Waldburger}}, \bibinfo
  {author} {\bibfnamefont {J.~G.}\ \bibnamefont {Danzl}}, \bibinfo {author}
  {\bibfnamefont {C.}~\bibnamefont {Chin}}, \bibinfo {author} {\bibfnamefont
  {B.}~\bibnamefont {Engeser}}, \bibinfo {author} {\bibfnamefont {A.~D.}\
  \bibnamefont {Lange}}, \bibinfo {author} {\bibfnamefont {K.}~\bibnamefont
  {Pilch}}, \bibinfo {author} {\bibfnamefont {A.}~\bibnamefont {Jaakkola}},
  \bibinfo {author} {\bibfnamefont {H.-C.}\ \bibnamefont {Naegerl}}, \ and\
  \bibinfo {author} {\bibfnamefont {R.}~\bibnamefont {Grimm}},\ }\href@noop {}
  {\bibfield  {journal} {\bibinfo  {journal} {Nature}\ }\textbf {\bibinfo
  {volume} {440}},\ \bibinfo {pages} {315} (\bibinfo {year}
  {2006})}\BibitemShut {NoStop}%
\bibitem [{\citenamefont {Gross}\ \emph {et~al.}(2009)\citenamefont {Gross},
  \citenamefont {Shotan}, \citenamefont {Kokkelmans},\ and\ \citenamefont
  {Khaykovich}}]{KhaykovichEfimov}%
  \BibitemOpen
  \bibfield  {author} {\bibinfo {author} {\bibfnamefont {N.}~\bibnamefont
  {Gross}}, \bibinfo {author} {\bibfnamefont {Z.}~\bibnamefont {Shotan}},
  \bibinfo {author} {\bibfnamefont {S.}~\bibnamefont {Kokkelmans}}, \ and\
  \bibinfo {author} {\bibfnamefont {L.}~\bibnamefont {Khaykovich}},\ }\href
  {\doibase 10.1103/PhysRevLett.103.163202} {\bibfield  {journal} {\bibinfo
  {journal} {Phys. Rev. Lett.}\ }\textbf {\bibinfo {volume} {103}},\ \bibinfo
  {pages} {163202} (\bibinfo {year} {2009})}\BibitemShut {NoStop}%
\bibitem [{\citenamefont {Gross}\ \emph {et~al.}(2010)\citenamefont {Gross},
  \citenamefont {Shotan}, \citenamefont {Kokkelmans},\ and\ \citenamefont
  {Khaykovich}}]{KhaykovichEfimov2}%
  \BibitemOpen
  \bibfield  {author} {\bibinfo {author} {\bibfnamefont {N.}~\bibnamefont
  {Gross}}, \bibinfo {author} {\bibfnamefont {Z.}~\bibnamefont {Shotan}},
  \bibinfo {author} {\bibfnamefont {S.}~\bibnamefont {Kokkelmans}}, \ and\
  \bibinfo {author} {\bibfnamefont {L.}~\bibnamefont {Khaykovich}},\
  }\href@noop {} {\bibfield  {journal} {\bibinfo  {journal} {Phys. Rev. Lett}\
  }\textbf {\bibinfo {volume} {105}},\ \bibinfo {pages} {103203} (\bibinfo
  {year} {2010})}\BibitemShut {NoStop}%
\bibitem [{\citenamefont {Berninger}\ \emph {et~al.}(2011)\citenamefont
  {Berninger}, \citenamefont {Zenesini}, \citenamefont {Huang}, \citenamefont
  {Harm}, \citenamefont {N\"{a}gerl}, \citenamefont {Ferlaino}, \citenamefont
  {Grimm}, \citenamefont {Julienne},\ and\ \citenamefont
  {Hutson}}]{Berninger2011uot}%
  \BibitemOpen
  \bibfield  {author} {\bibinfo {author} {\bibfnamefont {M.}~\bibnamefont
  {Berninger}}, \bibinfo {author} {\bibfnamefont {A.}~\bibnamefont {Zenesini}},
  \bibinfo {author} {\bibfnamefont {B.}~\bibnamefont {Huang}}, \bibinfo
  {author} {\bibfnamefont {W.}~\bibnamefont {Harm}}, \bibinfo {author}
  {\bibfnamefont {H.-C.}\ \bibnamefont {N\"{a}gerl}}, \bibinfo {author}
  {\bibfnamefont {F.}~\bibnamefont {Ferlaino}}, \bibinfo {author}
  {\bibfnamefont {R.}~\bibnamefont {Grimm}}, \bibinfo {author} {\bibfnamefont
  {P.~S.}\ \bibnamefont {Julienne}}, \ and\ \bibinfo {author} {\bibfnamefont
  {J.~M.}\ \bibnamefont {Hutson}},\ }\href@noop {} {\bibfield  {journal}
  {\bibinfo  {journal} {Phys. Rev. Lett.}\ }\textbf {\bibinfo {volume} {107}},\
  \bibinfo {pages} {120401} (\bibinfo {year} {2011})}\BibitemShut {NoStop}%
\bibitem [{\citenamefont {Wild}\ \emph {et~al.}(2012)\citenamefont {Wild},
  \citenamefont {Makotyn}, \citenamefont {Pino}, \citenamefont {Cornell},\ and\
  \citenamefont {Jin}}]{jin_contact_BEC}%
  \BibitemOpen
  \bibfield  {author} {\bibinfo {author} {\bibfnamefont {R.~J.}\ \bibnamefont
  {Wild}}, \bibinfo {author} {\bibfnamefont {P.}~\bibnamefont {Makotyn}},
  \bibinfo {author} {\bibfnamefont {J.~M.}\ \bibnamefont {Pino}}, \bibinfo
  {author} {\bibfnamefont {E.~A.}\ \bibnamefont {Cornell}}, \ and\ \bibinfo
  {author} {\bibfnamefont {D.~S.}\ \bibnamefont {Jin}},\ }\href@noop {}
  {\bibfield  {journal} {\bibinfo  {journal} {Phys. Rev. Lett.}\ }\textbf
  {\bibinfo {volume} {108}},\ \bibinfo {pages} {145305} (\bibinfo {year}
  {2012})}\BibitemShut {NoStop}%
\bibitem [{\citenamefont {Roy}\ \emph {et~al.}(2013)\citenamefont {Roy},
  \citenamefont {Landini}, \citenamefont {Trenkwalder}, \citenamefont
  {Semeghini}, \citenamefont {Spagnolli}, \citenamefont {Simoni}, \citenamefont
  {Fattori}, \citenamefont {Inguscio},\ and\ \citenamefont
  {Modugno}}]{Roy2013tot}%
  \BibitemOpen
  \bibfield  {author} {\bibinfo {author} {\bibfnamefont {S.}~\bibnamefont
  {Roy}}, \bibinfo {author} {\bibfnamefont {M.}~\bibnamefont {Landini}},
  \bibinfo {author} {\bibfnamefont {A.}~\bibnamefont {Trenkwalder}}, \bibinfo
  {author} {\bibfnamefont {G.}~\bibnamefont {Semeghini}}, \bibinfo {author}
  {\bibfnamefont {G.}~\bibnamefont {Spagnolli}}, \bibinfo {author}
  {\bibfnamefont {A.}~\bibnamefont {Simoni}}, \bibinfo {author} {\bibfnamefont
  {M.}~\bibnamefont {Fattori}}, \bibinfo {author} {\bibfnamefont
  {M.}~\bibnamefont {Inguscio}}, \ and\ \bibinfo {author} {\bibfnamefont
  {G.}~\bibnamefont {Modugno}},\ }\href {\doibase
  10.1103/PhysRevLett.111.053202} {\bibfield  {journal} {\bibinfo  {journal}
  {Phys. Rev. Lett.}\ }\textbf {\bibinfo {volume} {111}},\ \bibinfo {pages}
  {053202} (\bibinfo {year} {2013})}\BibitemShut {NoStop}%
\bibitem [{\citenamefont {Rem}\ \emph {et~al.}(2013)\citenamefont {Rem},
  \citenamefont {Grier}, \citenamefont {Ferrier-Barbut}, \citenamefont
  {Eismann}, \citenamefont {Langen}, \citenamefont {Navon}, \citenamefont
  {Khaykovich}, \citenamefont {Werner}, \citenamefont {Petrov}, \citenamefont
  {Chevy},\ and\ \citenamefont {Salomon}}]{UnitaryBoseGas}%
  \BibitemOpen
  \bibfield  {author} {\bibinfo {author} {\bibfnamefont {B.~S.}\ \bibnamefont
  {Rem}}, \bibinfo {author} {\bibfnamefont {A.~T.}\ \bibnamefont {Grier}},
  \bibinfo {author} {\bibfnamefont {I.}~\bibnamefont {Ferrier-Barbut}},
  \bibinfo {author} {\bibfnamefont {U.}~\bibnamefont {Eismann}}, \bibinfo
  {author} {\bibfnamefont {T.}~\bibnamefont {Langen}}, \bibinfo {author}
  {\bibfnamefont {N.}~\bibnamefont {Navon}}, \bibinfo {author} {\bibfnamefont
  {L.}~\bibnamefont {Khaykovich}}, \bibinfo {author} {\bibfnamefont
  {F.}~\bibnamefont {Werner}}, \bibinfo {author} {\bibfnamefont {D.~S.}\
  \bibnamefont {Petrov}}, \bibinfo {author} {\bibfnamefont {F.}~\bibnamefont
  {Chevy}}, \ and\ \bibinfo {author} {\bibfnamefont {C.}~\bibnamefont
  {Salomon}},\ }\href@noop {} {\bibfield  {journal} {\bibinfo  {journal} {Phys.
  Rev. Lett.}\ }\textbf {\bibinfo {volume} {110}},\ \bibinfo {pages} {163202}
  (\bibinfo {year} {2013})}\BibitemShut {NoStop}%
\bibitem [{\citenamefont {Fletcher}\ \emph {et~al.}(2013)\citenamefont
  {Fletcher}, \citenamefont {Gaunt}, \citenamefont {Navon}, \citenamefont
  {Smith},\ and\ \citenamefont {Hadzibabic}}]{hadzibabic_unitary_bose}%
  \BibitemOpen
  \bibfield  {author} {\bibinfo {author} {\bibfnamefont {R.~J.}\ \bibnamefont
  {Fletcher}}, \bibinfo {author} {\bibfnamefont {A.~L.}\ \bibnamefont {Gaunt}},
  \bibinfo {author} {\bibfnamefont {N.}~\bibnamefont {Navon}}, \bibinfo
  {author} {\bibfnamefont {R.~P.}\ \bibnamefont {Smith}}, \ and\ \bibinfo
  {author} {\bibfnamefont {Z.}~\bibnamefont {Hadzibabic}},\ }\href {\doibase
  10.1103/PhysRevLett.111.125303} {\bibfield  {journal} {\bibinfo  {journal}
  {Phys. Rev. Lett.}\ }\textbf {\bibinfo {volume} {111}},\ \bibinfo {pages}
  {125303} (\bibinfo {year} {2013})}\BibitemShut {NoStop}%
\bibitem [{\citenamefont {Huang}\ \emph
  {et~al.}(2014{\natexlab{a}})\citenamefont {Huang}, \citenamefont
  {Sidorenkov}, \citenamefont {Grimm},\ and\ \citenamefont
  {Hutson}}]{Huang2014oot}%
  \BibitemOpen
  \bibfield  {author} {\bibinfo {author} {\bibfnamefont {B.}~\bibnamefont
  {Huang}}, \bibinfo {author} {\bibfnamefont {L.~A.}\ \bibnamefont
  {Sidorenkov}}, \bibinfo {author} {\bibfnamefont {R.}~\bibnamefont {Grimm}}, \
  and\ \bibinfo {author} {\bibfnamefont {J.~M.}\ \bibnamefont {Hutson}},\
  }\href {\doibase 10.1103/PhysRevLett.112.190401} {\bibfield  {journal}
  {\bibinfo  {journal} {Phys. Rev. Lett.}\ }\textbf {\bibinfo {volume} {112}},\
  \bibinfo {pages} {190401} (\bibinfo {year} {2014}{\natexlab{a}})}\BibitemShut
  {NoStop}%
\bibitem [{Gri()}]{Grimm_finite_range_effects}%
  \BibitemOpen
  \href@noop {} {}\Eprint {http://arxiv.org/abs/B. Huang, L. A. Sidorenkov and
  R. Grimm, arXiv:1504.05360} {B. Huang, L. A. Sidorenkov and R. Grimm,
  arXiv:1504.05360} \BibitemShut {NoStop}%
\bibitem [{eis()}]{eismann_loss_dynamics}%
  \BibitemOpen
  \href@noop {} {}\Eprint {http://arxiv.org/abs/U. Eismann, L. Khaykovich, S.
  Laurent, I. Ferrier-Barbut, B. S. Rem, A. T. Grier, M. Delehaye, F. Chevy, C.
  Salomon, L.-C. Ha, C. Chin, arXiv:1505.04523} {U. Eismann, L. Khaykovich, S.
  Laurent, I. Ferrier-Barbut, B. S. Rem, A. T. Grier, M. Delehaye, F. Chevy, C.
  Salomon, L.-C. Ha, C. Chin, arXiv:1505.04523} \BibitemShut {NoStop}%
\bibitem [{\citenamefont {Ottenstein}\ \emph {et~al.}(2008)\citenamefont
  {Ottenstein}, \citenamefont {Lompe}, \citenamefont {Kohnen}, \citenamefont
  {Wenz},\ and\ \citenamefont {Jochim}}]{Ottenstein2008cso}%
  \BibitemOpen
  \bibfield  {author} {\bibinfo {author} {\bibfnamefont {T.~B.}\ \bibnamefont
  {Ottenstein}}, \bibinfo {author} {\bibfnamefont {T.}~\bibnamefont {Lompe}},
  \bibinfo {author} {\bibfnamefont {M.}~\bibnamefont {Kohnen}}, \bibinfo
  {author} {\bibfnamefont {A.~N.}\ \bibnamefont {Wenz}}, \ and\ \bibinfo
  {author} {\bibfnamefont {S.}~\bibnamefont {Jochim}},\ }\href {\doibase
  10.1103/PhysRevLett.101.203202} {\bibfield  {journal} {\bibinfo  {journal}
  {Phys. Rev. Lett.}\ }\textbf {\bibinfo {volume} {101}},\ \bibinfo {eid}
  {203202} (\bibinfo {year} {2008})}\BibitemShut {NoStop}%
\bibitem [{\citenamefont {Williams}\ \emph {et~al.}(2009)\citenamefont
  {Williams}, \citenamefont {Hazlett}, \citenamefont {Huckans}, \citenamefont
  {Stites}, \citenamefont {Zhang},\ and\ \citenamefont
  {O'Hara}}]{ohara_excited_trimer_li6}%
  \BibitemOpen
  \bibfield  {author} {\bibinfo {author} {\bibfnamefont {J.~R.}\ \bibnamefont
  {Williams}}, \bibinfo {author} {\bibfnamefont {E.~L.}\ \bibnamefont
  {Hazlett}}, \bibinfo {author} {\bibfnamefont {J.~H.}\ \bibnamefont
  {Huckans}}, \bibinfo {author} {\bibfnamefont {R.~W.}\ \bibnamefont {Stites}},
  \bibinfo {author} {\bibfnamefont {Y.}~\bibnamefont {Zhang}}, \ and\ \bibinfo
  {author} {\bibfnamefont {K.~M.}\ \bibnamefont {O'Hara}},\ }\href {\doibase
  10.1103/PhysRevLett.103.130404} {\bibfield  {journal} {\bibinfo  {journal}
  {Phys. Rev. Lett.}\ }\textbf {\bibinfo {volume} {103}},\ \bibinfo {pages}
  {130404} (\bibinfo {year} {2009})}\BibitemShut {NoStop}%
\bibitem [{\citenamefont {Huang}\ \emph
  {et~al.}(2014{\natexlab{b}})\citenamefont {Huang}, \citenamefont {O'Hara},
  \citenamefont {Grimm}, \citenamefont {Hutson},\ and\ \citenamefont
  {Petrov}}]{3BP_li6}%
  \BibitemOpen
  \bibfield  {author} {\bibinfo {author} {\bibfnamefont {B.}~\bibnamefont
  {Huang}}, \bibinfo {author} {\bibfnamefont {K.~M.}\ \bibnamefont {O'Hara}},
  \bibinfo {author} {\bibfnamefont {R.}~\bibnamefont {Grimm}}, \bibinfo
  {author} {\bibfnamefont {J.~M.}\ \bibnamefont {Hutson}}, \ and\ \bibinfo
  {author} {\bibfnamefont {D.~S.}\ \bibnamefont {Petrov}},\ }\href {\doibase
  10.1103/PhysRevA.90.043636} {\bibfield  {journal} {\bibinfo  {journal} {Phys.
  Rev. A}\ }\textbf {\bibinfo {volume} {90}},\ \bibinfo {pages} {043636}
  (\bibinfo {year} {2014}{\natexlab{b}})}\BibitemShut {NoStop}%
\bibitem [{\citenamefont {Barontini}\ \emph {et~al.}(2009)\citenamefont
  {Barontini}, \citenamefont {Weber}, \citenamefont {Rabatti}, \citenamefont
  {Catani}, \citenamefont {Thalhammer}, \citenamefont {Inguscio},\ and\
  \citenamefont {Minardi}}]{Florence_KRb}%
  \BibitemOpen
  \bibfield  {author} {\bibinfo {author} {\bibfnamefont {G.}~\bibnamefont
  {Barontini}}, \bibinfo {author} {\bibfnamefont {C.}~\bibnamefont {Weber}},
  \bibinfo {author} {\bibfnamefont {F.}~\bibnamefont {Rabatti}}, \bibinfo
  {author} {\bibfnamefont {J.}~\bibnamefont {Catani}}, \bibinfo {author}
  {\bibfnamefont {G.}~\bibnamefont {Thalhammer}}, \bibinfo {author}
  {\bibfnamefont {M.}~\bibnamefont {Inguscio}}, \ and\ \bibinfo {author}
  {\bibfnamefont {F.}~\bibnamefont {Minardi}},\ }\href {\doibase
  10.1103/PhysRevLett.103.043201} {\bibfield  {journal} {\bibinfo  {journal}
  {Phys. Rev. Lett.}\ }\textbf {\bibinfo {volume} {103}},\ \bibinfo {pages}
  {043201} (\bibinfo {year} {2009})}\BibitemShut {NoStop}%
\bibitem [{\citenamefont {Bloom}\ \emph {et~al.}(2013)\citenamefont {Bloom},
  \citenamefont {Hu}, \citenamefont {Cumby},\ and\ \citenamefont
  {Jin}}]{Bloom2013tou}%
  \BibitemOpen
  \bibfield  {author} {\bibinfo {author} {\bibfnamefont {R.~S.}\ \bibnamefont
  {Bloom}}, \bibinfo {author} {\bibfnamefont {M.-G.}\ \bibnamefont {Hu}},
  \bibinfo {author} {\bibfnamefont {T.~D.}\ \bibnamefont {Cumby}}, \ and\
  \bibinfo {author} {\bibfnamefont {D.~S.}\ \bibnamefont {Jin}},\ }\href
  {\doibase 10.1103/PhysRevLett.111.105301} {\bibfield  {journal} {\bibinfo
  {journal} {Phys. Rev. Lett.}\ }\textbf {\bibinfo {volume} {111}},\ \bibinfo
  {pages} {105301} (\bibinfo {year} {2013})}\BibitemShut {NoStop}%
\bibitem [{\citenamefont {Pires}\ \emph {et~al.}(2014)\citenamefont {Pires},
  \citenamefont {Ulmanis}, \citenamefont {H\"afner}, \citenamefont {Repp},
  \citenamefont {Arias}, \citenamefont {Kuhnle},\ and\ \citenamefont
  {Weidem\"uller}}]{heidelberg_LiCs_efimov}%
  \BibitemOpen
  \bibfield  {author} {\bibinfo {author} {\bibfnamefont {R.}~\bibnamefont
  {Pires}}, \bibinfo {author} {\bibfnamefont {J.}~\bibnamefont {Ulmanis}},
  \bibinfo {author} {\bibfnamefont {S.}~\bibnamefont {H\"afner}}, \bibinfo
  {author} {\bibfnamefont {M.}~\bibnamefont {Repp}}, \bibinfo {author}
  {\bibfnamefont {A.}~\bibnamefont {Arias}}, \bibinfo {author} {\bibfnamefont
  {E.~D.}\ \bibnamefont {Kuhnle}}, \ and\ \bibinfo {author} {\bibfnamefont
  {M.}~\bibnamefont {Weidem\"uller}},\ }\href {\doibase
  10.1103/PhysRevLett.112.250404} {\bibfield  {journal} {\bibinfo  {journal}
  {Phys. Rev. Lett.}\ }\textbf {\bibinfo {volume} {112}},\ \bibinfo {pages}
  {250404} (\bibinfo {year} {2014})}\BibitemShut {NoStop}%
\bibitem [{\citenamefont {Tung}\ \emph {et~al.}(2014)\citenamefont {Tung},
  \citenamefont {Jim\'enez-Garc\'ia}, \citenamefont {Johansen}, \citenamefont
  {Parker},\ and\ \citenamefont {Chin}}]{chin_LiCs_efimov}%
  \BibitemOpen
  \bibfield  {author} {\bibinfo {author} {\bibfnamefont {S.-K.}\ \bibnamefont
  {Tung}}, \bibinfo {author} {\bibfnamefont {K.}~\bibnamefont
  {Jim\'enez-Garc\'ia}}, \bibinfo {author} {\bibfnamefont {J.}~\bibnamefont
  {Johansen}}, \bibinfo {author} {\bibfnamefont {C.~V.}\ \bibnamefont
  {Parker}}, \ and\ \bibinfo {author} {\bibfnamefont {C.}~\bibnamefont
  {Chin}},\ }\href {\doibase 10.1103/PhysRevLett.113.240402} {\bibfield
  {journal} {\bibinfo  {journal} {Phys. Rev. Lett.}\ }\textbf {\bibinfo
  {volume} {113}},\ \bibinfo {pages} {240402} (\bibinfo {year}
  {2014})}\BibitemShut {NoStop}%
\bibitem [{\citenamefont {D'Incao}\ \emph {et~al.}(2004)\citenamefont
  {D'Incao}, \citenamefont {Suno},\ and\ \citenamefont {Esry}}]{DIncao2004lou}%
  \BibitemOpen
  \bibfield  {author} {\bibinfo {author} {\bibfnamefont {J.~P.}\ \bibnamefont
  {D'Incao}}, \bibinfo {author} {\bibfnamefont {H.}~\bibnamefont {Suno}}, \
  and\ \bibinfo {author} {\bibfnamefont {B.~D.}\ \bibnamefont {Esry}},\
  }\href@noop {} {\bibfield  {journal} {\bibinfo  {journal} {Phys. Rev. Lett.}\
  }\textbf {\bibinfo {volume} {93}},\ \bibinfo {pages} {123201} (\bibinfo
  {year} {2004})}\BibitemShut {NoStop}%
\bibitem [{\citenamefont {Efimov}(1970)}]{Efimov}%
  \BibitemOpen
  \bibfield  {author} {\bibinfo {author} {\bibfnamefont {V.~N.}\ \bibnamefont
  {Efimov}},\ }\href@noop {} {\bibfield  {journal} {\bibinfo  {journal} {Yad.
  Fiz.}\ }\textbf {\bibinfo {volume} {12}},\ \bibinfo {pages} {1080} (\bibinfo
  {year} {1970})},\ \bibinfo {note} {[Sov. J. Nucl. Phys. {\bf 12}, 589
  (1971)]}\BibitemShut {NoStop}%
\bibitem [{\citenamefont {Efimov}(1973)}]{Efimov73}%
  \BibitemOpen
  \bibfield  {author} {\bibinfo {author} {\bibfnamefont {V.}~\bibnamefont
  {Efimov}},\ }\href@noop {} {\bibfield  {journal} {\bibinfo  {journal} {Nucl.
  Phys.}\ }\textbf {\bibinfo {volume} {A 210}},\ \bibinfo {pages} {157}
  (\bibinfo {year} {1973})}\BibitemShut {NoStop}%
\bibitem [{\citenamefont {D'Incao}\ and\ \citenamefont
  {Esry}(2006)}]{DIncao2006eto}%
  \BibitemOpen
  \bibfield  {author} {\bibinfo {author} {\bibfnamefont {J.~P.}\ \bibnamefont
  {D'Incao}}\ and\ \bibinfo {author} {\bibfnamefont {B.~D.}\ \bibnamefont
  {Esry}},\ }\href@noop {} {\bibfield  {journal} {\bibinfo  {journal} {Phys.
  Rev. A}\ }\textbf {\bibinfo {volume} {73}},\ \bibinfo {pages} {030703(R)}
  (\bibinfo {year} {2006})}\BibitemShut {NoStop}%
\bibitem [{\citenamefont {Efimov}()}]{Efimov1979}%
  \BibitemOpen
  \bibfield  {author} {\bibinfo {author} {\bibfnamefont {V.}~\bibnamefont
  {Efimov}},\ }\href@noop {} {\bibinfo  {journal} {Yad. Fiz. {\bf 29}, {\rm
  1058 (1979) [}Sov. J. Nucl. Phys. {\bf 29}, {\rm 546 (1979)]}}\ }\BibitemShut
  {NoStop}%
\bibitem [{\citenamefont {Braaten}\ and\ \citenamefont
  {Hammer}(2006)}]{RevueBraaten}%
  \BibitemOpen
\bibfield  {journal} {  }\bibfield  {author} {\bibinfo {author} {\bibfnamefont
  {E.}~\bibnamefont {Braaten}}\ and\ \bibinfo {author} {\bibfnamefont {H.-W.}\
  \bibnamefont {Hammer}},\ }\href@noop {} {\bibfield  {journal} {\bibinfo
  {journal} {Phys. Rept.}\ }\textbf {\bibinfo {volume} {428}},\ \bibinfo
  {pages} {259} (\bibinfo {year} {2006})}\BibitemShut {NoStop}%
\bibitem [{\citenamefont {Braaten}\ \emph {et~al.}(2008)\citenamefont
  {Braaten}, \citenamefont {Hammer}, \citenamefont {Kang},\ and\ \citenamefont
  {Platter}}]{Braaten_rec_T}%
  \BibitemOpen
  \bibfield  {author} {\bibinfo {author} {\bibfnamefont {E.}~\bibnamefont
  {Braaten}}, \bibinfo {author} {\bibfnamefont {H.-W.}\ \bibnamefont {Hammer}},
  \bibinfo {author} {\bibfnamefont {D.}~\bibnamefont {Kang}}, \ and\ \bibinfo
  {author} {\bibfnamefont {L.}~\bibnamefont {Platter}},\ }\href@noop {}
  {\bibfield  {journal} {\bibinfo  {journal} {Phys. Rev. A}\ }\textbf {\bibinfo
  {volume} {78}},\ \bibinfo {pages} {043605} (\bibinfo {year}
  {2008})}\BibitemShut {NoStop}%
\bibitem [{\citenamefont {Marte}\ \emph {et~al.}(2002)\citenamefont {Marte},
  \citenamefont {Volz}, \citenamefont {Schuster}, \citenamefont {D\"urr},
  \citenamefont {Rempe}, \citenamefont {van Kempen},\ and\ \citenamefont
  {Verhaar}}]{Marte2002fri}%
  \BibitemOpen
  \bibfield  {author} {\bibinfo {author} {\bibfnamefont {A.}~\bibnamefont
  {Marte}}, \bibinfo {author} {\bibfnamefont {T.}~\bibnamefont {Volz}},
  \bibinfo {author} {\bibfnamefont {J.}~\bibnamefont {Schuster}}, \bibinfo
  {author} {\bibfnamefont {S.}~\bibnamefont {D\"urr}}, \bibinfo {author}
  {\bibfnamefont {G.}~\bibnamefont {Rempe}}, \bibinfo {author} {\bibfnamefont
  {E.~G.~M.}\ \bibnamefont {van Kempen}}, \ and\ \bibinfo {author}
  {\bibfnamefont {B.~J.}\ \bibnamefont {Verhaar}},\ }\href@noop {} {\bibfield
  {journal} {\bibinfo  {journal} {Phys. Rev. Lett.}\ }\textbf {\bibinfo
  {volume} {89}},\ \bibinfo {pages} {283202} (\bibinfo {year}
  {2002})}\BibitemShut {NoStop}%
\bibitem [{\citenamefont {Chin}\ \emph {et~al.}(2010)\citenamefont {Chin},
  \citenamefont {Grimm}, \citenamefont {Julienne},\ and\ \citenamefont
  {Tiesinga}}]{FeshbachRMP2010}%
  \BibitemOpen
  \bibfield  {author} {\bibinfo {author} {\bibfnamefont {C.}~\bibnamefont
  {Chin}}, \bibinfo {author} {\bibfnamefont {R.}~\bibnamefont {Grimm}},
  \bibinfo {author} {\bibfnamefont {P.}~\bibnamefont {Julienne}}, \ and\
  \bibinfo {author} {\bibfnamefont {E.}~\bibnamefont {Tiesinga}},\ }\href
  {\doibase 10.1103/RevModPhys.82.1225} {\bibfield  {journal} {\bibinfo
  {journal} {Rev. Mod. Phys.}\ }\textbf {\bibinfo {volume} {82}},\ \bibinfo
  {pages} {1225} (\bibinfo {year} {2010})}\BibitemShut {NoStop}%
\bibitem [{\citenamefont {Deh}\ \emph {et~al.}(2008)\citenamefont {Deh},
  \citenamefont {Marzok}, \citenamefont {Zimmermann},\ and\ \citenamefont
  {Courteille}}]{ZimmermannLi6RbRes}%
  \BibitemOpen
  \bibfield  {author} {\bibinfo {author} {\bibfnamefont {B.}~\bibnamefont
  {Deh}}, \bibinfo {author} {\bibfnamefont {C.}~\bibnamefont {Marzok}},
  \bibinfo {author} {\bibfnamefont {C.}~\bibnamefont {Zimmermann}}, \ and\
  \bibinfo {author} {\bibfnamefont {P.~W.}\ \bibnamefont {Courteille}},\ }\href
  {\doibase 10.1103/PhysRevA.77.010701} {\bibfield  {journal} {\bibinfo
  {journal} {Phys. Rev. A}\ }\textbf {\bibinfo {volume} {77}},\ \bibinfo
  {pages} {010701} (\bibinfo {year} {2008})}\BibitemShut {NoStop}%
\bibitem [{\citenamefont {Marzok}\ \emph {et~al.}(2009)\citenamefont {Marzok},
  \citenamefont {Deh}, \citenamefont {Zimmermann}, \citenamefont {Courteille},
  \citenamefont {Tiemann}, \citenamefont {Vanne},\ and\ \citenamefont
  {Saenz}}]{ZimmermannLi7RbRes}%
  \BibitemOpen
  \bibfield  {author} {\bibinfo {author} {\bibfnamefont {C.}~\bibnamefont
  {Marzok}}, \bibinfo {author} {\bibfnamefont {B.}~\bibnamefont {Deh}},
  \bibinfo {author} {\bibfnamefont {C.}~\bibnamefont {Zimmermann}}, \bibinfo
  {author} {\bibfnamefont {P.~W.}\ \bibnamefont {Courteille}}, \bibinfo
  {author} {\bibfnamefont {E.}~\bibnamefont {Tiemann}}, \bibinfo {author}
  {\bibfnamefont {Y.~V.}\ \bibnamefont {Vanne}}, \ and\ \bibinfo {author}
  {\bibfnamefont {A.}~\bibnamefont {Saenz}},\ }\href {\doibase
  10.1103/PhysRevA.79.012717} {\bibfield  {journal} {\bibinfo  {journal} {Phys.
  Rev. A}\ }\textbf {\bibinfo {volume} {79}},\ \bibinfo {pages} {012717}
  (\bibinfo {year} {2009})}\BibitemShut {NoStop}%
\bibitem [{\citenamefont {Petrov}\ \emph {et~al.}(2005)\citenamefont {Petrov},
  \citenamefont {Salomon},\ and\ \citenamefont {Shlyapnikov}}]{PetrovJPhysB}%
  \BibitemOpen
  \bibfield  {author} {\bibinfo {author} {\bibfnamefont {D.~S.}\ \bibnamefont
  {Petrov}}, \bibinfo {author} {\bibfnamefont {C.}~\bibnamefont {Salomon}}, \
  and\ \bibinfo {author} {\bibfnamefont {G.~V.}\ \bibnamefont {Shlyapnikov}},\
  }\href@noop {} {\bibfield  {journal} {\bibinfo  {journal} {J. Phys. B}\
  }\textbf {\bibinfo {volume} {38}},\ \bibinfo {pages} {S645} (\bibinfo {year}
  {2005})}\BibitemShut {NoStop}%
\bibitem [{\citenamefont {Helfrich}\ \emph {et~al.}(2010)\citenamefont
  {Helfrich}, \citenamefont {Hammer},\ and\ \citenamefont
  {Petrov}}]{HelfrichHammerPetrov}%
  \BibitemOpen
  \bibfield  {author} {\bibinfo {author} {\bibfnamefont {K.}~\bibnamefont
  {Helfrich}}, \bibinfo {author} {\bibfnamefont {H.-W.}\ \bibnamefont
  {Hammer}}, \ and\ \bibinfo {author} {\bibfnamefont {D.~S.}\ \bibnamefont
  {Petrov}},\ }\href@noop {} {\bibfield  {journal} {\bibinfo  {journal} {Phys.
  Rev. A}\ }\textbf {\bibinfo {volume} {81}},\ \bibinfo {pages} {042715}
  (\bibinfo {year} {2010})}\BibitemShut {NoStop}%
\bibitem [{\citenamefont {Berninger}\ \emph {et~al.}(2013)\citenamefont
  {Berninger}, \citenamefont {Zenesini}, \citenamefont {Huang}, \citenamefont
  {Harm}, \citenamefont {N\"agerl}, \citenamefont {Ferlaino}, \citenamefont
  {Grimm}, \citenamefont {Julienne},\ and\ \citenamefont
  {Hutson}}]{Berninger2013frw}%
  \BibitemOpen
  \bibfield  {author} {\bibinfo {author} {\bibfnamefont {M.}~\bibnamefont
  {Berninger}}, \bibinfo {author} {\bibfnamefont {A.}~\bibnamefont {Zenesini}},
  \bibinfo {author} {\bibfnamefont {B.}~\bibnamefont {Huang}}, \bibinfo
  {author} {\bibfnamefont {W.}~\bibnamefont {Harm}}, \bibinfo {author}
  {\bibfnamefont {H.-C.}\ \bibnamefont {N\"agerl}}, \bibinfo {author}
  {\bibfnamefont {F.}~\bibnamefont {Ferlaino}}, \bibinfo {author}
  {\bibfnamefont {R.}~\bibnamefont {Grimm}}, \bibinfo {author} {\bibfnamefont
  {P.~S.}\ \bibnamefont {Julienne}}, \ and\ \bibinfo {author} {\bibfnamefont
  {J.~M.}\ \bibnamefont {Hutson}},\ }\href {\doibase
  10.1103/PhysRevA.87.032517} {\bibfield  {journal} {\bibinfo  {journal} {Phys.
  Rev. A}\ }\textbf {\bibinfo {volume} {87}},\ \bibinfo {pages} {032517}
  (\bibinfo {year} {2013})}\BibitemShut {NoStop}%
\bibitem [{\citenamefont {Fonseca}\ \emph {et~al.}(1979)\citenamefont
  {Fonseca}, \citenamefont {Redish},\ and\ \citenamefont
  {Shanley}}]{FonsecaBO1979}%
  \BibitemOpen
  \bibfield  {author} {\bibinfo {author} {\bibfnamefont {A.~C.}\ \bibnamefont
  {Fonseca}}, \bibinfo {author} {\bibfnamefont {E.~F.}\ \bibnamefont {Redish}},
  \ and\ \bibinfo {author} {\bibfnamefont {P.~E.}\ \bibnamefont {Shanley}},\
  }\href@noop {} {\bibfield  {journal} {\bibinfo  {journal} {Nucl. Phys. A}\
  }\textbf {\bibinfo {volume} {320}},\ \bibinfo {pages} {273} (\bibinfo {year}
  {1979})}\BibitemShut {NoStop}%
\bibitem [{\citenamefont {Yamashita}\ \emph {et~al.}(2013)\citenamefont
  {Yamashita}, \citenamefont {Bellotti}, \citenamefont {Frederico},
  \citenamefont {Fedorov}, \citenamefont {Jensen},\ and\ \citenamefont
  {Zinner}}]{Yamashita2013spm}%
  \BibitemOpen
  \bibfield  {author} {\bibinfo {author} {\bibfnamefont {M.~T.}\ \bibnamefont
  {Yamashita}}, \bibinfo {author} {\bibfnamefont {F.~F.}\ \bibnamefont
  {Bellotti}}, \bibinfo {author} {\bibfnamefont {T.}~\bibnamefont {Frederico}},
  \bibinfo {author} {\bibfnamefont {D.~V.}\ \bibnamefont {Fedorov}}, \bibinfo
  {author} {\bibfnamefont {A.~S.}\ \bibnamefont {Jensen}}, \ and\ \bibinfo
  {author} {\bibfnamefont {N.~T.}\ \bibnamefont {Zinner}},\ }\href@noop {}
  {\bibfield  {journal} {\bibinfo  {journal} {Phys. Rev. A}\ }\textbf {\bibinfo
  {volume} {87}},\ \bibinfo {pages} {062702} (\bibinfo {year}
  {2013})}\BibitemShut {NoStop}%
\bibitem [{\citenamefont {Ulmanis}\ \emph {et~al.}(2015)\citenamefont
  {Ulmanis}, \citenamefont {H\"afner}, \citenamefont {Pires}, \citenamefont
  {Kuhnle}, \citenamefont {Weidem\"uller},\ and\ \citenamefont
  {Tiemann}}]{heidelberg_LiCs_Fesh_Res}%
  \BibitemOpen
  \bibfield  {author} {\bibinfo {author} {\bibfnamefont {J.}~\bibnamefont
  {Ulmanis}}, \bibinfo {author} {\bibfnamefont {S.}~\bibnamefont {H\"afner}},
  \bibinfo {author} {\bibfnamefont {R.}~\bibnamefont {Pires}}, \bibinfo
  {author} {\bibfnamefont {E.~D.}\ \bibnamefont {Kuhnle}}, \bibinfo {author}
  {\bibfnamefont {M.}~\bibnamefont {Weidem\"uller}}, \ and\ \bibinfo {author}
  {\bibfnamefont {E.}~\bibnamefont {Tiemann}},\ }\href@noop {} {\bibfield
  {journal} {\bibinfo  {journal} {New J. Phys.}\ }\textbf {\bibinfo {volume}
  {17}},\ \bibinfo {pages} {055009} (\bibinfo {year} {2015})}\BibitemShut
  {NoStop}%
\bibitem [{\citenamefont {Derevianko}\ \emph {et~al.}(2001)\citenamefont
  {Derevianko}, \citenamefont {Babb},\ and\ \citenamefont
  {Dalgarno}}]{c6_hetero}%
  \BibitemOpen
  \bibfield  {author} {\bibinfo {author} {\bibfnamefont {A.}~\bibnamefont
  {Derevianko}}, \bibinfo {author} {\bibfnamefont {J.~F.}\ \bibnamefont
  {Babb}}, \ and\ \bibinfo {author} {\bibfnamefont {A.}~\bibnamefont
  {Dalgarno}},\ }\href {\doibase 10.1103/PhysRevA.63.052704} {\bibfield
  {journal} {\bibinfo  {journal} {Phys. Rev. A}\ }\textbf {\bibinfo {volume}
  {63}},\ \bibinfo {pages} {052704} (\bibinfo {year} {2001})}\BibitemShut
  {NoStop}%
\bibitem [{Pet()}]{PetrovLesHouches2010}%
  \BibitemOpen
  \href@noop {} {}\bibinfo {note} {D.~S.~Petrov, in {\it Proceedings of the Les
  Houches Summer Schools, Session 94}, edited by C. Salomon, G. V. Shlyapnikov,
  and L. F. Cugliandolo (Oxford University Press, Oxford, England, 2013),
  e-print arXiv:1206.5752.}\BibitemShut {Stop}%
\bibitem [{\citenamefont {Tung}\ \emph {et~al.}(2013)\citenamefont {Tung},
  \citenamefont {Parker}, \citenamefont {Johansen}, \citenamefont {Chin},
  \citenamefont {Wang},\ and\ \citenamefont
  {Julienne}}]{chin_LiCs_Feshb_Res_2013}%
  \BibitemOpen
  \bibfield  {author} {\bibinfo {author} {\bibfnamefont {S.-K.}\ \bibnamefont
  {Tung}}, \bibinfo {author} {\bibfnamefont {C.}~\bibnamefont {Parker}},
  \bibinfo {author} {\bibfnamefont {J.}~\bibnamefont {Johansen}}, \bibinfo
  {author} {\bibfnamefont {C.}~\bibnamefont {Chin}}, \bibinfo {author}
  {\bibfnamefont {Y.}~\bibnamefont {Wang}}, \ and\ \bibinfo {author}
  {\bibfnamefont {P.~S.}\ \bibnamefont {Julienne}},\ }\href {\doibase
  10.1103/PhysRevA.87.010702} {\bibfield  {journal} {\bibinfo  {journal} {Phys.
  Rev. A}\ }\textbf {\bibinfo {volume} {87}},\ \bibinfo {pages} {010702}
  (\bibinfo {year} {2013})}\BibitemShut {NoStop}%
\bibitem [{\citenamefont {Castin}\ and\ \citenamefont {Werner}()}]{LeChapitre}%
  \BibitemOpen
  \bibfield  {author} {\bibinfo {author} {\bibfnamefont {Y.}~\bibnamefont
  {Castin}}\ and\ \bibinfo {author} {\bibfnamefont {F.}~\bibnamefont
  {Werner}},\ }\href@noop {} {}\Eprint {http://arxiv.org/abs/in {\sl The
  BCS-BEC Crossover and the Unitary Fermi Gas}, Lecture Notes in Physics {\bf
  836}, 127, W. Zwerger ed. (Springer, Heidelberg, 2012)} {in {\sl The BCS-BEC
  Crossover and the Unitary Fermi Gas}, Lecture Notes in Physics {\bf 836},
  127, W. Zwerger ed. (Springer, Heidelberg, 2012)} \BibitemShut {NoStop}%
\bibitem [{\citenamefont {Wang}\ \emph {et~al.}(2012)\citenamefont {Wang},
  \citenamefont {Wang}, \citenamefont {D'Incao},\ and\ \citenamefont
  {Greene}}]{Wang2012utb}%
  \BibitemOpen
  \bibfield  {author} {\bibinfo {author} {\bibfnamefont {Y.}~\bibnamefont
  {Wang}}, \bibinfo {author} {\bibfnamefont {J.}~\bibnamefont {Wang}}, \bibinfo
  {author} {\bibfnamefont {J.~P.}\ \bibnamefont {D'Incao}}, \ and\ \bibinfo
  {author} {\bibfnamefont {C.~H.}\ \bibnamefont {Greene}},\ }\href@noop {}
  {\bibfield  {journal} {\bibinfo  {journal} {Phys. Rev. Lett.}\ }\textbf
  {\bibinfo {volume} {109}},\ \bibinfo {pages} {243201} (\bibinfo {year}
  {2012})}\BibitemShut {NoStop}%
\bibitem [{\citenamefont {Wang}\ and\ \citenamefont
  {Esry}(2011)}]{WangNJP2011}%
  \BibitemOpen
  \bibfield  {author} {\bibinfo {author} {\bibfnamefont {Y.}~\bibnamefont
  {Wang}}\ and\ \bibinfo {author} {\bibfnamefont {B.~D.}\ \bibnamefont
  {Esry}},\ }\href {\doibase 10.1088/1367-2630/13/3/035025} {\bibfield
  {journal} {\bibinfo  {journal} {New J. Phys.}\ }\textbf {\bibinfo {volume}
  {13}},\ \bibinfo {pages} {035025} (\bibinfo {year} {2011})}\BibitemShut
  {NoStop}%
\bibitem [{\citenamefont {Laurent}\ \emph {et~al.}(2014)\citenamefont
  {Laurent}, \citenamefont {Leyronas},\ and\ \citenamefont
  {Chevy}}]{chevy_nk_with_losses}%
  \BibitemOpen
  \bibfield  {author} {\bibinfo {author} {\bibfnamefont {S.}~\bibnamefont
  {Laurent}}, \bibinfo {author} {\bibfnamefont {X.}~\bibnamefont {Leyronas}}, \
  and\ \bibinfo {author} {\bibfnamefont {F.}~\bibnamefont {Chevy}},\ }\href
  {\doibase 10.1103/PhysRevLett.113.220601} {\bibfield  {journal} {\bibinfo
  {journal} {Phys. Rev. Lett.}\ }\textbf {\bibinfo {volume} {113}},\ \bibinfo
  {pages} {220601} (\bibinfo {year} {2014})}\BibitemShut {NoStop}%
\bibitem [{\citenamefont {Petrov}(2003)}]{Petrov3fermions}%
  \BibitemOpen
  \bibfield  {author} {\bibinfo {author} {\bibfnamefont {D.~S.}\ \bibnamefont
  {Petrov}},\ }\href@noop {} {\bibfield  {journal} {\bibinfo  {journal} {Phys.
  Rev. A}\ }\textbf {\bibinfo {volume} {67}},\ \bibinfo {pages} {010703}
  (\bibinfo {year} {2003})}\BibitemShut {NoStop}%
\bibitem [{\citenamefont {Castin}\ and\ \citenamefont
  {Tignone}(2011)}]{castin_tignone}%
  \BibitemOpen
  \bibfield  {author} {\bibinfo {author} {\bibfnamefont {Y.}~\bibnamefont
  {Castin}}\ and\ \bibinfo {author} {\bibfnamefont {E.}~\bibnamefont
  {Tignone}},\ }\href {\doibase 10.1103/PhysRevA.84.062704} {\bibfield
  {journal} {\bibinfo  {journal} {Phys. Rev. A}\ }\textbf {\bibinfo {volume}
  {84}},\ \bibinfo {pages} {062704} (\bibinfo {year} {2011})}\BibitemShut
  {NoStop}%
\bibitem [{\citenamefont {Macek}\ \emph {et~al.}(2005)\citenamefont {Macek},
  \citenamefont {Ovchinnikov},\ and\ \citenamefont {Gasaneo}}]{Macek1}%
  \BibitemOpen
  \bibfield  {author} {\bibinfo {author} {\bibfnamefont {J.~H.}\ \bibnamefont
  {Macek}}, \bibinfo {author} {\bibfnamefont {S.}~\bibnamefont {Ovchinnikov}},
  \ and\ \bibinfo {author} {\bibfnamefont {G.}~\bibnamefont {Gasaneo}},\
  }\href@noop {} {\bibfield  {journal} {\bibinfo  {journal} {Phys. Rev. A}\
  }\textbf {\bibinfo {volume} {72}},\ \bibinfo {pages} {032709} (\bibinfo
  {year} {2005})}\BibitemShut {NoStop}%
\bibitem [{\citenamefont {Macek}\ \emph {et~al.}(2006)\citenamefont {Macek},
  \citenamefont {Ovchinnikov},\ and\ \citenamefont {Gasaneo}}]{Macek2}%
  \BibitemOpen
  \bibfield  {author} {\bibinfo {author} {\bibfnamefont {J.~H.}\ \bibnamefont
  {Macek}}, \bibinfo {author} {\bibfnamefont {S.~Y.}\ \bibnamefont
  {Ovchinnikov}}, \ and\ \bibinfo {author} {\bibfnamefont {G.}~\bibnamefont
  {Gasaneo}},\ }\href@noop {} {\bibfield  {journal} {\bibinfo  {journal} {Phys.
  Rev. A}\ }\textbf {\bibinfo {volume} {73}},\ \bibinfo {pages} {032704}
  (\bibinfo {year} {2006})}\BibitemShut {NoStop}%
\bibitem [{\citenamefont {Gogolin}\ \emph {et~al.}(2008)\citenamefont
  {Gogolin}, \citenamefont {Mora},\ and\ \citenamefont
  {Egger}}]{Gogolin2008aso}%
  \BibitemOpen
  \bibfield  {author} {\bibinfo {author} {\bibfnamefont {A.~O.}\ \bibnamefont
  {Gogolin}}, \bibinfo {author} {\bibfnamefont {C.}~\bibnamefont {Mora}}, \
  and\ \bibinfo {author} {\bibfnamefont {R.}~\bibnamefont {Egger}},\
  }\href@noop {} {\bibfield  {journal} {\bibinfo  {journal} {Phys. Rev. Lett.}\
  }\textbf {\bibinfo {volume} {100}},\ \bibinfo {pages} {140404} (\bibinfo
  {year} {2008})}\BibitemShut {NoStop}%
\bibitem [{\citenamefont {Mora}\ \emph {et~al.}(2011)\citenamefont {Mora},
  \citenamefont {Gogolin},\ and\ \citenamefont {Egger}}]{MoraCRAS}%
  \BibitemOpen
  \bibfield  {author} {\bibinfo {author} {\bibfnamefont {C.}~\bibnamefont
  {Mora}}, \bibinfo {author} {\bibfnamefont {A.~O.}\ \bibnamefont {Gogolin}}, \
  and\ \bibinfo {author} {\bibfnamefont {R.}~\bibnamefont {Egger}},\
  }\href@noop {} {\bibfield  {journal} {\bibinfo  {journal} {C. R. Physique}\
  }\textbf {\bibinfo {volume} {12}},\ \bibinfo {pages} {27} (\bibinfo {year}
  {2011})}\BibitemShut {NoStop}%
\bibitem [{\citenamefont {Mikkelsen}\ \emph {et~al.}(2015)\citenamefont
  {Mikkelsen}, \citenamefont {Jensen}, \citenamefont {Fedorov},\ and\
  \citenamefont {Zinner}}]{ZinnerHetero}%
  \BibitemOpen
  \bibfield  {author} {\bibinfo {author} {\bibfnamefont {M.}~\bibnamefont
  {Mikkelsen}}, \bibinfo {author} {\bibfnamefont {A.~S.}\ \bibnamefont
  {Jensen}}, \bibinfo {author} {\bibfnamefont {D.~V.}\ \bibnamefont {Fedorov}},
  \ and\ \bibinfo {author} {\bibfnamefont {N.~T.}\ \bibnamefont {Zinner}},\
  }\href@noop {} {\bibfield  {journal} {\bibinfo  {journal} {J. Phys. B}\
  }\textbf {\bibinfo {volume} {48}},\ \bibinfo {pages} {085301} (\bibinfo
  {year} {2015})}\BibitemShut {NoStop}%
\bibitem [{\citenamefont {Castin}\ and\ \citenamefont
  {Werner}(2011)}]{CastinWerner_nk_trimer}%
  \BibitemOpen
  \bibfield  {author} {\bibinfo {author} {\bibfnamefont {Y.}~\bibnamefont
  {Castin}}\ and\ \bibinfo {author} {\bibfnamefont {F.}~\bibnamefont
  {Werner}},\ }\href@noop {} {\bibfield  {journal} {\bibinfo  {journal} {Phys.
  Rev. A}\ }\textbf {\bibinfo {volume} {83}},\ \bibinfo {pages} {063614}
  (\bibinfo {year} {2011})}\BibitemShut {NoStop}%
\bibitem [{\citenamefont {Gradshteyn}\ and\ \citenamefont
  {Ryzhik}(1994)}]{Gradstein}%
  \BibitemOpen
  \bibfield  {author} {\bibinfo {author} {\bibfnamefont {I.~S.}\ \bibnamefont
  {Gradshteyn}}\ and\ \bibinfo {author} {\bibfnamefont {I.~M.}\ \bibnamefont
  {Ryzhik}},\ }\href@noop {} {\emph {\bibinfo {title} {Tables of integrals,
  series, and products}}}\ (\bibinfo  {publisher} {Academic Press},\ \bibinfo
  {year} {1994})\ \bibinfo {note} {5th Edition, A. Jeffrey, Editor}\BibitemShut
  {NoStop}%
\bibitem [{\citenamefont {Werner}\ and\ \citenamefont
  {Castin}(2012)}]{WernerCastinRelationsBosons}%
  \BibitemOpen
  \bibfield  {author} {\bibinfo {author} {\bibfnamefont {F.}~\bibnamefont
  {Werner}}\ and\ \bibinfo {author} {\bibfnamefont {Y.}~\bibnamefont
  {Castin}},\ }\href@noop {} {\bibfield  {journal} {\bibinfo  {journal} {Phys.
  Rev. A}\ }\textbf {\bibinfo {volume} {86}},\ \bibinfo {pages} {053633}
  (\bibinfo {year} {2012})}\BibitemShut {NoStop}%
\end{thebibliography}%

\end{document}